\documentclass[12pt]{iopart}
\usepackage{epsfig}
\usepackage{graphicx}
\usepackage{lscape}
\usepackage{longtable}
\usepackage{deluxetable}

\usepackage{iopams}  
\begin{document}

\title[Youthful Nearby Binary System]{Accretion and OH Photodissociation at a Nearby T Tauri System in the $\beta$ Pictoris Moving Group}

\author{B. Zuckerman$^1$, Laura Vican$^1$, and David R. Rodriguez$^2$}

\address{$^1$Department of Physics and Astronomy, University of California, Los Angeles, CA 90095, USA}
\address{$^2$Departamento de Astronomia, Universidad de Chile, Casilla 36-D, Santiago, Chile}
\eads{\mailto{ben@astro.ucla.edu}, \mailto{lvican@ucla.edu}, \mailto{drodrigu@das.uchile.cl}}
\begin{abstract}
We present spectra of an M-type, binary star system (LDS 5606) that belongs to the nearby $\sim$20 Myr old $\beta$ Pictoris moving group.  Both stars are very dusty; the dustier member displays optical emission lines from eight elements indicative of ongoing mass accretion.   The spectra of both stars contain oxygen forbidden line emission at 6302 and 5579 \AA, consistent with a recent model of far ultraviolet photodissociation of OH molecules in a circumstellar disk.  These are the oldest dwarf stars presently known to display such a phenomenon.  The spectral energy distribution of the dustier star indicates substantial quantities of dust as hot as 900 K, and its fractional infrared luminosity (L$_{IR}$/L$_{bol}$) is almost as large as that of the main sequence record holder, V488 Per.   The LDS 5606 binary joins a remarkable group of very dusty, old, T Tauri stars that belong to widely separated multiple systems.
\end{abstract}
\maketitle

\section{INTRODUCTION}

Beginning in the late 1990s, astronomers have identified numerous youthful stars near Earth (see reviews by Zuckerman \& Song 2004, Torres et al 2008, Malo et al 2013, Gagne et al 2014).  Some of these stars belong to associations comprised of members that are moving, more or less, together through space.   The principal tool for identification of such stars less massive than mid F-type has been the ROSAT all-sky X-ray survey (RASS).   Recently it was realized that many late-type dwarf stars with excess ultraviolet emission in the GALEX sky survey (Martin et al 2005) are both young and within $\sim$100 pc of Earth (Rodriguez et al 2011, 2013; Shkolnik et al 2011).  Indeed, in comparison with the RASS, the GALEX survey provides a more sensitive means for identification of youthful M-type stars.

As part of an ongoing survey of GALEX UV-excess stars, Rodriguez et al (2014) identified an unusual binary star, LDS 5606, that they show is a likely member of the nearby $\beta$ Pictoris moving group.  Following previous designations in the literature, we designate the west and east members of the binary as LDS 5606A and LDS 5606B, respectively, with a projected separation on the sky of 26".  At a distance of 65 pc (Rodriguez et al 2014), this corresponds to $\sim$1700 AU.  The age of the $\beta$ Pic stars has been deduced via placement on evolutionary tracks (Zuckerman et al 2001), kinematic traceback (Ortega et al 2002), and lithium depletion (Binks \& Jeffries 2014 and references therein); these ages range from 12 to 21 Myr.   At present, lithium depletion ages for youthful stars seem to be very much in favor.  Thus we adopt a $\beta$ Pic moving group age of 20 Myr.

At an age as old as 20 Myr, both the dust and the gas that orbit the two LDS 5606 stars display unusual properties.  Rodriguez et al (2014) note that for a binary star system of this age and older the presence of detectable dust in orbit about both members is very unusual.  Rodriguez et al establish that both members of LDS 5606 are of spectral type $\sim$M5.  We know of no (published) evidence for the presence of orbiting gas around late-type dwarf stars with ages $\geq$20 Myr, but a few such dusty main sequence stars of spectral types G and A are orbited by substantial masses of gas.  The gas has been detected optically, in the far-infrared, and at millimeter wavelengths; the following paragraph presents a brief overview.   

The optical CaII H and K lines are seen toward two members of the $\beta$ Pictoris moving group, $\beta$ Pic itself and HD 172555 (Kiefer et al 2014 and references therein).  Such gas is usually interpreted in a model of falling evaporating bodies and the mass of gas implied in such a model is far smaller than the mass of gas implied by detection of the [OI] 63 $\mu$m line or by rotational emission from the CO molecule.   With the Herschel Space Observatory the 63 $\mu$m [O I] line has been detected in emission toward HD 172555 (Riviere-Marichalar et al 2012) and the 158 $\mu$m [CII] line toward $\beta$ Pic (Cataldi et al 2014).  CO rotational emission has been seen toward
V4046 Sgr (Rodriguez et al 2010), 49 Cet (Zuckerman \& Song 2012), HD 21997 (Kospal et al 2013), and $\beta$ Pic (Dent et al 2014) with ages in the range 20-40 Myr.    V4046 Sgr -- a member of the $\beta$ Pic moving group (Torres et al 2008) -- shows evidence for accretion (Stempels \& Gahm 2004), but not of hot dust (Hutchinson et al 1990; Jensen \& Mathieu 1997; Rosenfeld et al 2013).   Because V 4046 Sgr is a 2.4-day binary system, the accretion is likely to be driven by complex gas dynamics induced by the close binarity (de Val-Borro et al 2011; Sytov et al 2011; Donati et al 2011).  HD 21997 and 49 Cet show only cold dust and no evidence for accretion.

The oldest stars -- not in a close binary and not yet evolved beyond the main sequence -- previously known to be undergoing gas accretion reside in the 8 Myr old TW Association, for example TWA 30AB (Looper et al 2010a, 2010b) and TW Hya itself (Kastner et al 2002; Curran et al 2011).  As shown in Section 4, the dustier (west) star of the LDS 5606 binary is definitely accreting material from a surrounding disk.  Also present in the spectra of both members of the binary are forbidden lines of neutral oxygen at vacuum wavelengths of 6302 and 5579 \AA.   As we describe in Section 4.4, these lines are plausibly an outcome of photodissociation by far-ultraviolet photons of OH molecules located in disks that orbit each star.   Such spectral features are unprecedented in dwarf stars of age 20 Myr (or older).

\section{OBSERVATIONS}

The spectra discussed in the present paper were obtained primarily with the HIRES echelle spectrometer (Vogt et al. 1994) on the Keck I telescope at Mauna Kea Observatory in Hawaii.  Spectra were obtained during observing runs in October and November 2013.  The red cross disperser was combined with a 1.15'' slit and the wavelength range between 4370 and  9000 \AA\ was covered with a resolution of $\sim$40,000.   HIRES employs three CCDs with spectral gaps between them and between some orders as listed in Table A1.  Spectra were reduced using the IRAF and MAKEE software packages.  Representative spectra appear in Figures 1-6.   All wavelengths given in this paper are in vacuum.
A ThAr lamp and velocity standards HIP 26689, HIP 47690, and HIP 117473 (Nidever et al 2002) were used to determine heliocentric radial velocities.

Given their nearly identical JHK magnitudes in the Two Micron All Sky Survey (2MASS; Skrutskie et al 2006), their excess emission at UV and IR wavelengths, strong veiling in the western (A) star, and the large errors in optical magnitudes listed in VizieR, neither component of the binary star can be convincingly designated as the (brighter) primary.    LDS 5606A was observed on 2013 October 17 and November 16 (UT).  The echelle grating settings were not the same in October and November so the wavelength coverage differed at the extreme wavelengths and also in the gaps listed in Table A1.  This is why some of the transitions listed in Tables 1 and 2 for LDS 5606A have only a single entry.   Clouds were present during both nights and hampered the integrations (which were terminated early). Integration times were 1900 and 1713 sec on 10/17 and 11/16, respectively.

The eastern component, (B), was observed on 21 October 2013 (UT).   Some thin clouds were present on this night.  The integration time was 2300 sec.  HIRES results for LDS 5606B are presented in Tables 3 and 4.  

In Tables 1-4, equivalent widths (EW) and line center wavelengths were obtained by fitting lines with Voigt profiles using the IRAF function splot.    In a very few cases the lines have irregular shapes and the Voigt fits are quite poor.  In such cases EWs were measured by summing the area either above (for emission lines) or below (for absorption lines) the continuum level.  Examples are the He I emission line at 4389 \AA\ and the questionable Sc I absorption line at 4739 \AA\ as measured on 10/17/13 in LDS 5606A.  Line full widths at half maximum intensity (FWHM), discussed in Section 4, are obtained by fitting Gaussian profiles.

We used Voigt profiles in order to capture the full EW of a line.  It was obvious for some of the strong lines with good signal-to-noise ratio (S/N) that a Gaussian fit missed energy in the linewings and thus would underestimate the EW.  However at half-power both the Gaussian and the Lorentzian components of the Voigt profile contribute to the FWHM in a Voigt fit.  This is unnecessarily complex for our needs; the Gaussian FWHM dominates and is sufficient for inter- comparison of linewidths (as per Section 4).

In Sections 3 and 4 we also utilize data obtained on 2013 December 3 (UT) with the UVES spectrograph (Dekker et al 2000) on the Very Large Telescope (VLT).   Details are given in Rodriguez et al (2014); for the present paper we note that spectra were obtained from 3760 to 5000 \AA, from 5700 to 7500 \AA, and from 7660 to 9460 \AA\ with a 2.1" slit and a spectral resolution $\sim$20,000.  The integration time was only 900 sec per star so that the S/N in the UVES spectra is inferior to that of HIRES.  Tables 5 and 6 and Figures 6-8 present some UVES results.  For lines in the blue the UVES measured continuum level is so faint that EW measurements have large uncertainties.  Fortunately, as part of the UVES pipeline, appropriate calibrations were taken on December 3 so that the blue portion of the two UVES spectra were flux calibrated.  Thus, Table 5 lists emission line flux densities rather than EW.  At longer wavelengths, Table 6 presents some measured EW from the UVES spectra.

In September and November 2013 both members of the binary star were observed with the APEX telescope (G\"usten et al 2006) at the J = 3$\rightarrow$2 rotational transition of CO with spectral resolution 0.53 km s$^{-1}$ per channel.  Water vapor columns ranged between 0.3 and 4.2 mm.   A total of 248.8 minutes were spent on LDS 5606A and 140.7 minutes on LDS 5606B.  The summed spectra for the A (western) star had an rms in antenna temperature (per 0.53 km s$^{-1}$ channel) of 7.17 mK and for the B (eastern) star 5.47 mK.

\section{RESULTS}

Much basic information about LDS 5606A and 5606B is given in Table 1 of Rodriguez et al (2014),  for example their spectral types (M5), distance from Earth ($\sim$65 pc), and radial velocities.  A full discussion of the radial velocities can be found in Section 2 of that paper.

Following Vican \& Schneider (2014), spectral energy distributions
(SEDs) were created with a fully automated fitting technique that uses theoretical models from Hauschildt et al. (1999) to predict stellar photospheric fluxes.  The SEDs were generated using available photometry from Hipparcos, Tycho-2, 2MASS, and ALLWISE (Wright et al 2014).  Stellar radii and effective temperatures are treated as free parameters to fit the observed fluxes (B, V, J, H, and K) with a $\chi$$^{2}$ minimization method. 

The SEDs for the two stars are shown in Figures 9 to 11.   
Their fractional infrared luminosities (L$_{IR}$/L$_{bol}$, 11.9 and 6.3\%) are among the largest known for any dwarf star with age $>$10 Myr.   
More than half of the infrared luminosity of LDS 5606A is carried by dust particles with temperatures $\sim$900 K. 
Excepting much more modest quantities of hot dust seen interferometrically at K-band by Absil et al (2013) around some much older, earlier type (A $-$ K) stars,  this is the hottest dust ever seen at a dwarf star of age $>$10 Myr.
Previous to the measurements of LDS 5606A, the hottest dust measured at a dwarf star far from interstellar molecular clouds and with properties similar to the dust at LDS 5606A is $\sim$800 K at the $\sim$60 Myr old K-type star V488 Per (see Section 4.6.2 and Zuckerman et al 2012).   

As indicated in Table 1, emission lines from at least 8 elements are seen in the spectrum of LDS 5606A.   For most lines the identification of the carrier seems secure; where the carrier is less certain we have added a (?) after what appears to be the most plausible ion.   Many marginal (low S/N, $\sim$5) emission features appear in the spectra.   Future, deeper, integrations will clarify the reality of such features and may well add additional elements in emission.

Comparison of the two epochs of observation of LDS 5606A with each other and with the one epoch observation of LDS 5606B indicates the following:

1. The numerous and strong permitted emission lines at the western (A)  star combined with the large full width at half maximum intensity (FWHM) of the H$\alpha$ line indicate an accretion flow (see Section 4).  The weaker and fewer emission lines at the eastern (B)  star can probably be accounted for by some combination of an active chromosphere and accretion.  The difference in emission line intensities between the two stars is nicely illustrated in the broadband UVES spectra in Figure 3 in Rodriguez et al (2014).

2.  The presence of 900 K dust (Figure 9) near LDS 5606A is consistent with the ongoing accretion implied by the optical emission lines.   If blackbodies, then these dust particles lie
only $\sim$0.01 AU from LDS 5606A -- similar to the semimajor axes of some
Kepler discovered planets, and just where one might expect to find lots of
dust if 20 Myr old planetary embryos of such close-in planets are colliding with each other (e.g. Hansen \& Murray 2013).
 
3.  The EW of the permitted emission lines from LDS 5606A were typically about 18\% stronger on 10/17 than on 11/16 (Table 1).   H$\alpha$ is a notable exception, being substantially stronger on 11/16.   No overall change is evident in the average absorption line EWs between the two epochs.   

4.  To determine the veiling in the optical spectra of the two stars we compared the EW of their absorption lines with those of a star of similar spectral type but with no evidence for excess IR emission and much weaker emission lines (see Section 4.1).  The conclusion is that LDS 5606A is heavily veiled and LDS 5606B modestly so (Table 7).

5.  [OI] lines at 6302 and 5579 \AA\ at both stars (Figure 6) indicate the presence of relatively low-density atomic gas; we attribute the [OI] to far ultraviolet photodissociation of OH molecules located in circumstellar disks (Section 4.4).

\section{Discussion}

\subsection{Veiling}

The presence of overlying (excess) optical continuum emission at classical T Tauri stars is referred to as veiling.   In the popular magnetospheric model, this emission can be generated by dissipation of energy of infalling  gas in a postshock region at the base of a magnetic funnel at the stellar surface  (Bertout 1989; Beristain et al 2001 and references therein).    The shocked region is also a plausible strong source of ionizing radiation and/or a high temperature region, essential for excitation of the helium lines observed in LDS 5606A (Section 4.3).

White \& Basri (2003) characterize the excess continuum emission (excess flux density) with a quantity r, defined as r = F$_{excess}$/F$_{photosphere}$.  They measure r by comparing the depths of photospheric absorption 
features in the spectrum of a T Tauri star with the depths of the same features in the spectrum of a standard star.   


To determine the veiling of LDS 5606A and B we compare their spectra with that of a (likely) unveiled star of similar spectral type chosen from our HIRES survey of UV-bright stars (L. Vican et al, in preparation). 
The star we use, 2MASS 05294468-3239141, was studied by Riaz et al (2006) and classified as M4.5.  Based on its K$_s$ minus W3 color, the star would be classified as M5 according to Table 6 in Pecaut \& Mamajek (2013).  2M0529-32 has few and only modest emission lines in our HIRES spectrum and likely little or no accretion luminosity.  In this respect and also in its spectral type it is similar to four other stars from our HIRES survey whose SEDs are all plotted in Figure 11.

Based on the relative strengths of a few dozen absorption lines in the HIRES spectra of 2M0529-32 and the LDS 5606 stars we calculate the veiling (r values) listed in Table 7.   According to Section 4.2 of White \& Basri (2003), stars with r $>$0.06 are veiled and are classical T Tauri stars.   Thus, according to this criterion and our Table 7, both components of LDS 5606 are veiled.

\subsection{Accretion}

Figure 7 in White \& Basri (2003) is a plot of the EW of the H$\alpha$ line vs its 10\% width in various T Tauri stars.  In Figure 12 we overplot on the White \& Basri figure the HIRES measurements of H$\alpha$ in the two LDS 5606 stars.   Based on Figure 12, LDS 5606A is accreting material.   Figure 3 in Natta et al (2004) displays accretion rate vs the H$\alpha$ 10\% width for stars and brown dwarfs with masses $<$0.23 $M_{\odot}$; this mass range includes the LDS 5606 stars ($\sim$0.1 $M_{\odot}$, Rodriguez et al 2014).   According to the Natta et al figure, the LDS 5606A mass accretion rate probably lies in the range 10$^{-10}$ to 10$^{-11}$ $M_{\odot}$ yr$^{-1}$.

Other evidence for accretion in LDS 5606A are veiling (Section 4.1 and Table 7) and also the many emission lines (Table 1) not normally seen in the spectra of young active stars.  For example in a sample of $\sim$40 UV-bright late-type stars observed with HIRES (L. Vican et al, in preparation), only one shows emission lines beyond those from H, He, the Na I doublet near 5900 \AA, and the Ca II IR triplet.  Of note at LDS 5606A is the strong high excitation Paschen $\alpha$ line of ionized helium at 4687 \AA, indicative of the existence of a region either of high temperature or close proximity to a source of ionizing radiation (Section 4.3).  This line is not present in the spectra of young non-accreting stars (Vican et al, in preparation).

The situation regarding accretion at LDS 5606B is more ambiguous.  We base our analysis on the discussion in White \& Basri (2003) who consider three different indicators for accretion; these are veiling, H$\alpha$ EW, and the 10\% width of the H$\alpha$ line.  According to Section 3.3 in White \& Basri, veiling of a star's optical continuum is evidence for accretion and LDS 54606B is veiled (Section 4.1 and Table 7).   According to White \& Basri's Section 4.2, an M5-type T Tauri star is "classical", i.e. accreting, if its H$\alpha$ EW is $\geq$20 \AA, which is true for LDS 5606B at least some of the time, but not always (see Table 2 in Rodriguez et al 2014).
On the other hand, the third accretion indicator, the H$\alpha$ 10\% width in LDS 5606B is only $\sim$135 km s$^{-1}$ (Figure 12) which places the star squarely in the region of weak-line (non-accreting) stars according to the discussion in White \& Basri.

Thus, LDS 5606B appears to pass the first and maybe the second of these three tests for accretion, but fails the third, thus placing it in a region of ``phase space'' not occupied by any of the much younger T Tauri stars investigated by White \& Basri (2003).  The referee suggests that these data might be reconciled if accretion at LDS 5606B is sporadic in nature.

An additional indicator that LDS 5606B actively accretes at least some of the time is its unusually strong near-UV flux measured by GALEX.  As noted by Rodriguez et al (2014), the near-UV emission from LDS 5606B was so strong that it fell outside of the standard region in color-color space occupied by candidate young late-type stars as defined in Rodriguez et al (2013), a situation reminiscent of the actively-accreting young star TW Hya (Rodriguez et al 2011). 

\subsection{H and He Emission Lines}

As noted in Section 4.2, the EW and 10\% width of the H$\alpha$ emission lines in the LDS 5606 stars can be used to deduce mass accretion rates.  
Out beyond the 10\% width, the extent of the extreme velocities of H and He emission lines can be used to better understand physical conditions in the various mass flows near T Taurs stars, as discussed in the following paragraphs.  We therefore consider the full velocity extent (FVE) of lines in LDS 5606.  Our definition of FVE is that velocity at which the line wings appear to merge with the continuum.  Because we are interested only in comparing FVEs among various transitions and stars -- where the FVE differences among the transitions and stars are substantial (see below) -- 
for the purposes of the present paper, FVE need not be defined too precisely.  

The FVE of the
H$\alpha$ and H$\beta$ lines in LDS 5606A are much greater than the corresponding lines in LDS 5606B (Fig 1).   For H$\alpha$, the FVE is $\sim$870 km s$^{-1}$ in the A-componant compared to $\sim$550 km s$^{-1}$ in B.   At H$\beta$ the difference between the two stars is greater; the FVE in A is $\sim$1370 km s$^{-1}$, while for B the FVE is $\sim$550 km s$^{-1}$.   

Muzerolle et al (2001) attribute broad H$\alpha$ line wings to Stark broadening of energy levels of H atoms that are participating in the funneled flow of material onto a T Tauri star in the magnetospheric model.   In contrast, Beristain et al (2001) propose that in some T Tauri stars H$\alpha$ emission at velocity offsets more than 200 km s$^{-1}$ from the stellar velocity is generated in a hot wind.  They suggest that a choice can be made between these two models for high-velocity gas by measurement of asymmetry, if any, in the red and blue high velocity wings; a modest red asymmetry would favor the infall model and a blue asymmetry the outflow model.  However, no convincing asymmetry is   present in either the H$\alpha$ or H$\beta$ line wings in LDS 5606A.  The infall model rather than the existence of a substantial, rapidly outflowing, wind is consistent with the generally modest FWHM of the emission lines in LDS 5606A.  The broadest FWHM of any line is H$\alpha$ at 138 km s$^{-1}$ followed by H$\beta$ at 123 km s$^{-1}$.   The broadest helium lines have FWHM $\sim$50 km s$^{-1}$, but more typically $\sim$25 km s$^{-1}$, while for heavier elements emission FWHM are typically 20 km s$^{-1}$ or less.   In classical T Tauri stars rapid outflow is characterized by much broader emission lines (Beristain et al 2001).

Table 1 lists seven detected transitions of He I and one of He II in LDS 5606A.\footnote{Three additional transitions of He I, at vacuum wavelengths of 4438.8, 5049.1, and 7283.3 \AA, are covered in the HIRES spectra, but none are detected.   A He I transition at 7067.1 \AA\ falls in a hole in the HIRES  coverage (see Table A1), but is detected in the UVES spectrum of LDS 5606A (Table 6).} Because of the high energies of the He energy levels, their excitation is generally attributed to photoionization and subsequent recombination and cascade rather than collisional excitation which would require very high gas temperatures.   In the magnetospheric model the accretion shock at the stellar surface is deemed to be the principal source of ionizing photons.   Beristain et al (2001) decompose the He lines from classical T Tauri stars into two kinematic components, narrow and broad ($\sim$50 and $\sim$200 km s$^{-1}$, respectively, see their Table 1).  The narrow component arises in conjunction with postshock gas at the magnetospheric footprint while the broad component arises preferentially in a hot outflowing wind.   The relatively narrow linewidths mentioned in the preceding paragraph in combination with the predominance of permitted emission lines rather than forbidden lines (see Sections 4.4 and 4.6.1) indicate that the spectrum of LDS 5606A is dominated by the narrow component.  Thus, in the following we compare helium line ratios in LDS 5606A with those arising from the narrow component in classical T Tauri stars as analyzed by Beristain et al (2001).

Beristain et al (2001) focus their attention on the ratio of the intensity of the 5878 \AA\ line of ortho-He to that of the 6680 \AA\ line of para-He.   The ratios of these intensities have been worked out for nebular (HII region) conditions ranging from pure capture and cascade in the low-density limit to much higher densities where collisions dominate (see discussion and references in Beristain et al).   Like Beristain et al, our HIRES spectra are not flux calibrated so, like them, we use the ratio of EWs of the two lines.   Over the modest 800 \AA\ separation of the two lines, the combination of photospheric and veiled continuum at LDS 5606A is probably not much different from unity.  


The 5878/6680 EW ratio is 4.7 (Table 1), consistent with densities $\sim$10,000 cm$^{-3}$ and temperatures $\sim$20,000 K (Smits 1991) where collisional excitation of the metastable triplet and singlet 2S states can compete with radiative processes.  According to the calculations by Smits (1991), a ratio of 4.7 is near the maximum achievable at any nebular density.
The ratio of 4.7 is larger than those quoted by Beristain et al (2001) for the narrow helium line component in 24 classical T Tauri stars and implies densities in the emitting region at LDS 5606A smaller than those in younger T Tauri stars.   
For the 4473/4714 ratio we measure an EW ratio of 9.8, in agreement with the nebular model prediction of 9.8 (Smits 1991).

Turning to a joint consideration of the seven lines of He I and the 4687 \AA\ line of He II in LDS 5606A, their kinematic properties are consistent with origin in postshock gas near the base of the accretion flows in the magnetospheric model.  In Table 8 we compare the stellar velocity as deduced from strong photospheric absorption lines with the redshifts of the neutral and ionized He emission lines.  A progression of He I moderately redshifted relative to the photosphere, to He II redshifted relative to He I -- as measured in LDS 5606A, see Notes to Table 8 -- is consistent with association of helium emission lines and the accretion shock (Beristain et al (2001).  The larger FWHM of the 4687 \AA\ He II line (51 km s$^{-1}$), relative to the average FWHM of the seven He I lines (28.4 $\pm$9.2 km s$^{-1}$), is also anticipated in the magnetospheric model; the He II line would be formed closer to the accretion shock, where gas velocities, ionizing flux, and temperature are all relatively larger.   (These FWHM are from Gaussian fits to the emission lines.  The shape and width of the 4389 He I line are anomalous and there may be contamination from other elements.  If this line is removed from the calculation then the FWHM of the other 6 He I lines is 25.2$\pm$4.0 km s$^{-1}$.)  Thus the relative He linewidths and radial velocities are consistent with expectations of the magnetospheric shock model.

In the case where the helium lines result entirely from recombination and cascade, then the 4473/4687 intensity ratio would equal 0.04(N[HeII]/N[HeIII]) (Osterbrock 1974).   The measured EW ratio from Table 1 is 3.6.   The near coincidence in wavelength between the Lyman $\alpha$ transition in hydrogen and the Balmer $\beta$ (n = 4-2) transition in He+ provides a mechanism to increase the population of the He+ n=4 level, thus enhancing the intensity of the 4687 line.  Hence, the 4473/4687 ratio provides only a lower limit on the He II to He III number ratio.

\subsection{[O I] lines}

Tables 1, 3, and 6 and Figure 6 include the [O I] electric quadruple transition at 5579 \AA\ and the magnetic dipole transition at 6302 \AA.   In gaseous nebulae the former line, sometimes referred to as an ''auroral'' transition, is between the singlet S and singlet D term of the ground configuration.  The 6302 \AA\ line, a ''nebular'' transition in gaseous nebulae, is between the singlet D and the ground state triplet P$_2$ term.   There also is a magnetic dipole transition at 6365.5 \AA\ from the singlet D to the triplet P$_1$ term with Einstein A-value 1/3 that of the 6302 \AA\ transition.   The 6365.5 \AA\ line is not detected in the HIRES spectra; the EW upper limit is $\sim$120 m\AA\ in both members of LDS 5606.   

Forbidden line emission at classical T Tauri stars can come in two flavors, a high-velocity, strongly blueshifted component that comes from collimated outflowing jets and a low-velocity component with an, at most, modest blueshift.  The 5579 \AA\ line often accompanies the low-velocity component, but is never associated with the high-velocity one (Gorti et al 2011).  Various models to explain the origin of low-velocity [O I] emission at classical T Tauri stars have been proposed (Gorti et al 2011 and Rigliaco et al 2013 and references therein).   In the following we consider the HIRES [O I] measurements in the context of the OH photodissociation model proposed in these two papers.   According to this model, stellar far-UV photons reach the surface of an orbiting molecular disk where they photodissociate both H$_2$O and OH molecules, resulting in O atoms in the $^1$D and $^1$S states.  These atoms then decay via the 6302 and 5579 \AA\ transitions.  Since [O I] emission is present at both members of LDS 5606 and because forbidden line radiation is quenched at high densities (such as might characterize the region where the 900 K dust is located around LDS 5606A), it seems reasonable to presume that the [O I] lines are associated with the 200 K dust component that likely is located $\sim$0.34 AU from each star (see  Section 4.5).

An absolute minimum requirement for this model to work is that the far-UV luminosity be sufficient to produce the observed [O I] luminosity.   To account for a given 6302 \AA\ [O I] luminosity, Rigliaco et al (2013) estimate that the far-UV flux must be at least 20 times the [O I] flux and find (their Equation 3) that the average ratio of far-UV to [O I] luminosity in their sample of T Tauri stars is $\sim$2300 times this lower limit.   For LDS 5606A we use the [O I] EW given in Table 1 and a (rather uncertain) R-band magnitude of 15.2 (see entries in VizieR) for the continuum to obtain the [O I] luminosity.   The combination of the GALEX far-UV and near-UV channels Ð from 1344 to 2616 \AA\ (the photodissociation wavelength of OH) -- yields a Òfar-UVÓ luminosity that is $\sim$550 times the [O I] luminosity.   The effective far-UV luminosity for OH photodissociation by Lyman $\alpha$ photons could be an order of magnitude larger than the GALEX measured Òfar-UVÓ luminosity (Figure 1 in Bergin et al 2003).

The 6302/5579 EW ratio of 1.6$\pm$0.4 in LDS 5606A is at the low end of the range (1-8) of ratios for classical T Tauri stars given in Rigliaco et al (2013).  Because the HIRES spectra are not flux calibrated, conversion of the EW ratio to a ratio of luminosities is uncertain.  The photospheric blackbody continuum flux density at 6302 \AA\ would be about twice that at 5579 \AA, while according to Table 7 the veiling continuum might be slightly greater at 5579 \AA\ compared to 6302 \AA.  On net then, at LDS 5606A the luminosity of the 6302 \AA\ line would be about three times that of the 5579 \AA\ line.  According to the detailed discussion of the dust/gas disk around TW Hya in Gorti et al (2011), in the OH photodissociation model such a ratio is anticipated in dense disk gas where the collisional reaction  O($^1$D) + H$_2$ $\rightarrow$ OH  can compete with OH photodissociation rates.  If the 200 K dust/gas disk around LDS 5606A is similar to that around TW Hya, then the total gas number density in the [O I] emission region at LDS 5606A might be $\sim$10$^{10}$ cm$^{-3}$.  But more data and a disk model tailored to LDS 5606A are needed before any definitive conclusions can be drawn.

Typically the [O I] lines in T Tauri stars are slightly blueshifted with respect to the stellar velocity (Table 4 in Rigliaco et al 2013).  For LDS 5606A there is no blueshift, but perhaps a small redshift.  The [O I] linewidths at LDS 5606A are definitely at the low end of the range of FWHM of the T Tauri stars listed in Table 4 of Rigliaco et al (2013).   For the 6302 \AA\ line in LDS 5606A, with a Gaussian fit, we measure 14.3 km s$^{-1}$ and for the 5579 \AA\ line 8.4 km/s; these linewidths have been corrected for instrumental broadening.   The stars in Table 4 of Rigliaco et al. that best match the lack of blueshift and small width of the [O I] lines in LDS 5606A are TW Hya and DR Tau whose orbiting disks have small (7$^\circ$) and moderate (37$^\circ$) inclination angles, respectively, with respect to the line of sight.   Perhaps the disk at LDS 5606A also has a small inclination angle so that we are observing it nearly face-on.   The contrast between the emission line patterns seen in LDS 5606A and TWA30 (described in Section 4.6.1), would be consistent with a near face-on disk at LDS 5606A.  The apparent correlation between FWHM and inclination angle in the entries in Table 4 of Rogliaco et al (2013) suggest that disk inclination is a significant factor contributing to the relative [O I] linewidths seen among classical T Tauri stars.

\subsection{Dust and gas abundances}

In this Section we estimate a minimum mass of dust around each member of LDS 5606 by 
assuming thin shells of small particles at distances from the stars set by 
their SEDs (Figures 9 and 10).  For LDS 5606A the 900 K and 215 K dust 
particles absorb 7.2\% and 4.7\% of the starlight, respectively.  For blackbody particles the  215 K dust is about 18 times further from the star than the 900 K dust, so the minimum 
total mass of the cooler particles is about 300 times that of the hotter.  Thus, the 
calculations to follow pertain to the $\sim$220 K dust at both LDS 5606A and 
B and we refer to the star simply as LDS 5606.

We first estimate how far from LDS 5606 220 K blackbody dust particles would lie.  The effective temperature and luminosity of LDS 5606 are about 2880 K and 0.012 $L_{\odot}$(Rodriguez et al 2014).  Then the radius of LDS 5606 is $\sim$0.44 $R_{\odot}$.  The orbital semi-major axes (R$_d$)$_{BB}$ of blackbody particles is given by 

(R$_d$)$_{BB}$ = (R$_*$/2)(T$_*$/T$_d$)$^2$

where R$_*$ and T$_*$ are the radius and effective temperature of the central star and T$_d$ is the particle temperature at (R$_d$)$_{BB}$.  With the above T$_*$ and R$_*$ for LDS 5606 and 220 K dust particles, then (R$_d$)$_{BB}$ = 0.17 AU.

Figure 9 in Rodriguez \& Zuckerman (2012) indicates that for thermally emitting dust particles seen at debris disks, the true particulate semi-major axis, R$_d$, is typically about twice (R$_d$)$_{BB}$, or for LDS 5606, $\sim$0.34 AU. 
 
A minimum dust mass can be estimated by assuming that the entire infrared luminosity is due to absorption of starlight by grains of radius ÒaÓ equal to $\lambda_{peak}$ divided by 2$\pi$, where $\lambda_{peak}$ is the wavelength at peak stellar emission.  For LDS 5606, $\lambda_{peak}$ is about one $\mu$m and thus a is $\sim$0.2 $\mu$m.  At this optimum size, particles absorb starlight approximately as blackbodies; if a is substantially larger or smaller, then particle opacity per unit mass will be smaller. 

The minimum dust mass (M$_{min}$) required to account for the luminosity of 220 K dust particles is

M$_{min}$ = (16$\pi$/3)($\tau$)($\rho$)(a)(R$_d$)$^2$   

where $\rho$ is the density of an individual grain (e.g., Chen \& Jura 2001) and $\tau$ = L$_{IR}$/L$_{bol}$ is the infrared luminosity.  With $\rho$ = 2.5 g cm$^{-3}$, M$_{min}$ $\sim$10$^{21}$ g.  By comparison, the mass of the largest asteroid, Ceres, is $\sim$10$^{24}$ g.

There is no evidence for dust colder than 200 K and indeed such dust may not exist since it often does not for young stars that are members of wide binary systems (e.g., HD 15407, Melis et al 2010).  Therefore, we first estimate whether it would be possible for APEX to have detected optical thick J = 3-2 CO emission from a region around LDS 5606 with radius 0.34 AU.  That is, we assume coexisting gas and dust at a common temperature of 220 K.  A face-on disk of radius 0.34 AU at 65 pc will subtend a diameter of 0.01'' to be compared to the APEX beam diameter of 17''.  The beam averaged CO brightness temperature would then be  $\sim$0.1 mK, far smaller than the measured rms values of 7.2 and 5.5 mK for the western and eastern stars (see Section 2).
Thus, the regions around the two stars that are known to contain dust would be invisible to APEX even if CO J = 3-2 emission is optically thick.

Optimistically, one might assume a dust extent at LDS 5606 analogous to that of the young K8-type  star TW Hya.  The bolometric luminosity of TW Hya, 53 pc from Earth, is $\sim$0.26 $L_{\odot}$ (Ducourant et al 2014).  The radius of the CO emission region at TW Hya is $\sim$5'' (Rosenfeld et al 2012).  Scaling by the relative parallaxes and bolometric luminosities, one could thus anticipate a CO emission region at LDS 5606 of radius $\sim$0.8'', and a CO brightness temperature smaller by a factor of (0.8/5.0)$^2$ than that at TW Hya.
The peak CO J = 3-2 flux density at TW Hya measured with APEX is $\sim$30 Jy (J. Kastner, personal communication 2014).  With a conversion factor of 41 Jy/K, the TW Hya CO 3-2 brightness temperature is 0.7 K, yielding a potential measured CO temperature of $\sim$20 mK at LDS 5606; this is about 3 times the rms given in the previous paragraph.  Thus, about all we can say at this point is that if a disk at LDS 5606 exists far beyond the region that contains 220 K dust, then the LDS 5606 disk is not much richer in CO than is the one that orbits TW Hya.

\subsection{Comparison of LDS 5606 with other young, dusty, nearby stars}

In the present section we consider some young dusty stars that display some important similarities and differences with LDS 5606.

\subsubsection{TWA 30}

The visual binary star TWA 30 (Looper et al 2010a \& b) bears some striking similarities and differences in comparison to LDS 5606.    Both systems are composed of stars of similar spectral types ($\sim$M5) and with large projected separations in the plane of the sky, $\sim$1700 AU for LDS 5606 and $\sim$3400 AU for TWA 30.   TWA 30 is a member of the $\sim$8 Myr old TW Hydrae Association while LDS 5606 is a member of the $\sim$20 Myr old $\beta$ Pictoris moving group.    Unlike LDS 5606A, neither TWA 30A nor 30B appear in the GALEX catalog.

The optical-infrared spectral energy distributions (SEDs) of the less dusty of the two stars in each system are remarkably similar.  TWA 30A has 210 K dust with an infrared luminosity $\tau$ = 2.4\% (Schneider et al 2012), while the dust at LDS 5606B has a temperature of 220 K and $\tau$ = 6.3\% (Figure 10).   The dust emission at TWA 30B can be decomposed into two components of 190 and 660 K (Schneider et al 2012; Looper et al 2010b), similar to our dust temperature decomposition at LDS 5606A of 215 and 900 K (Figure 9).   The principal difference in the SEDs of TWA 30B and LDS 5606A is that the optical light in the former is largely extinguished by the nearly edge-on dust disk (Looper et al 2010b; Schneider et al 2012).

The optical spectra of TWA 30A and 30B and of LDS 5606A contain many emission lines, but only a few lines are common to the TWA stars and LDS 5606A.  The spectra of both TWA stars are dominated by forbidden emission lines whereas permitted emission lines characterize the LDS 5606A spectrum.   The only forbidden lines in the LDS 5606A spectrum are the two [O I] lines discussed in Section 4.4; their FWHM $\sim$10 km s$^{-1}$, are much narrower than the [O I] 6302 \AA\ lines in both TWA 30A and 30B ($\sim$100 km s$^{-1}$).    TWA 30A displays very weak He I 5877 \AA\ emission and TWA 30B displays no optical He line emission at all, in contrast to the many and strong helium lines in LDS 5606A.

Looper et al (2010a and b) interpret the TWA 30 spectra as due to a disk seen edge-on in the case of TWA 30B and nearly edge-on in the case of TWA 30A.  The many and broad forbidden lines in the optical spectra of both stars are produced in strong outflows nearly in the plane of the sky.   The edge-on disks obscure the regions near the base of the accretion flows, regions that are responsible for the strong permitted lines expected in the magnetospheric model and seen in many T Tauri stars.   By contrast, as mentioned in Section 4.4, the orientation of the dusty disk at LDS 5606A is more likely to be nearly face-on than edge-on.   This enables the optical helium and hydrogen lines to be seen.  In addition, the general absence of forbidden line emission indicates that any outflow at LDS 5606A is much weaker than at the TWA 30 stars.   As discussed in Section 4.4, the two oxygen forbidden lines detected in both LDS 5606 stars are likely formed by photodissociation of OH in its circumstellar disk and probably not in an outflow.

The relative X-ray properties of TWA 30 and LDS 5606 are considered in Section 3.3 of Rodriguez et al (2014).

\subsubsection{V488 Per}

V488 Per is a K-type member of the 60 Myr old $\alpha$ Persei cluster.  It is surrounded by a dusty disk with infrared SED similar to that of LDS 5606A (Zuckerman et al 2012).   The dust at V488 Per can be decomposed into components of temperatures 120 and 820 K with $\tau$ = 16\%.   These values are similar to those of LDS 5606A, 215 and 900 K and $\tau$ = 11.9\% (Figure 9).  Along with LDS 5606B, with $\tau$ = 6.3\% (Figure 10), these three stars have the largest fractional infrared luminosities of any known dwarf stars with ages $>$10 Myr.  V488 Per is probably a member of a wide triple system (B. Zuckerman, in preparation).

Unlike LDS 5606A, and notwithstanding the 820 K dust, the optical spectrum of V488 Per is quite bland.  The H$\alpha$ line EW -- measured from a high-quality HIRES spectrum obtained on UT 1997 December 12 by Dr. I. N. Reid and available from the Keck Observatory Archive -- is 280 m\AA\ in absorption.  The spectrum covers the range between 6340 and 8725 \AA; the only emission lines to be seen are the self-reversed cores of the Ca IR triplet.   Thus the physical conditions near the two stars are quite different.    Zuckerman et al (2012) suggest that the hot dust seen at V488 Per is a result of a collision of two planetary embryos.

The SEDs out to 22 $\mu$m at TWA 30B, V488 Per, and LDS 5606A can all be decomposed into two distinct temperature dust components, one quite hot and the other of moderate temperature.   The SEDs at wavelengths longer than $\sim$22$\mu$m are not known.   This common decomposition is suggestive of the existence of substantial planets with temperatures similar to Earth in orbit around all three stars.

\subsubsection{V4046 Sgr}

V4046 Sgr is a 2.4 day period binary composed of two K-type stars.   Like LDS 5606, it is a member of the $\beta$ Pic moving group (Torres et al 2008).   The SED shows no evidence for dust hotter than about 200 K (Hutchinson et al 1990; Jensen \& Mathieu 1997; Rosenfeld et al 2013).  Nonetheless, with the UVES echelle spectrometer on the VLT, Stempels \& Gahm (2004) detected strong emission lines from the Balmer series of hydrogen up to at least H16 and also the Ca II H- and K-lines, as well as  weak emission lines of He I at 5877 \AA\ and of [O I] at 6302 \AA.   They deduced the presence of a weak veiling continuum that contributes about 1/3 of the total observed flux between 3500 and 4000 \AA, decreasing to about 10\% of the total flux between 6000 and 6700 \AA.   Thus, some accretion appears to be ongoing, but because of complex dynamics induced by the close binarity (de Val-Borro et al 2011; Sytov et al 2011; Donati et al 2011), it is hard to compare this system with LDS 5606A.

\section{CONCLUSIONS}

We have identified an unusual low-mass binary star, LDS 5606, that Rodriguez et al (2014) classify as a member of the $\beta$ Pictoris moving group.   The western member of the system (LDS 5606A) is still actively accreting material from an orbiting dusty disk $\sim$20 Myr after its central star began to form $-$ this is more than twice as old as the oldest low mass stars previously known with measurable accretion signatures.  These include both members of the 8 Myr old binary star TWA 30 (Looper et al 2010a and b) that appear both remarkably similar to and yet different from LDS 5606A.

All four members of these two wide-separation binary stars are among the most infrared luminous dwarf stars (in the sense of large L$_{IR}$/L$_{bol}$) currently known in astronomy.   LDS 5606 thus can be added to a remarkable collection of nearby ''old'' and very dusty T Tauri type and more massive stars that are members of wide (of order 1000 to 10000 AU) multiple star systems.  In addition to LDS 5606 and TWA 30, these include TW Hya, Hen 3-600, HD 98800, HD 15407, V4046 Sgr, HR 4796, T Cha (Kastner et al 2012), and probably V488 Per.   The infrared spectral energy distribution (SED) of individual members of at least three of these systems (LDS 5606A, TWA 30B and V488 Per) can be fit with dust at two distinct temperatures -- one quite hot (700 -- 900 K) and the other cooler (100 -- 200 K) -- as if the dust is divided by planets forming in a region with temperature similar to that of Earth.

In their Conclusion section Kastner et al (2012) speculate about time scales for planet formation in hierarchical multiple systems and whether the presence of a wide binary or tertiary companion can have profound effects on planet building processes in close proximity to one or multiple members of such systems.  The ensemble of such ÒoldÓ systems now known may be sufficiently large that calculations of Kozai-driven evolution might be revealing.

$\smallskip$

We thank Dr. Adam Schneider for calculating the SEDs shown in Figures 9 \& 10 and Dr. R. White for supplying the data for Figure 7 of White \& Basri (2003).  We thank the referee for a timely and conscientious report.
This research was supported in part by a NASA ADAP grant to the Rochester Institute of Technology and to UCLA, and by an NSF pre-doctoral fellowship to Laura Vican.  D.R.R. acknowledges support from FONDECYT grant 3130520.
This paper includes data gathered with the Very Large Telescope under ESO
program ID 092.CÐ0203(A). Most of the 
data presented herein were obtained at the W.M. Keck Observatory, which is operated as a scientific partnership among the California Institute of Technology, the University of California, and the National Aeronautics and Space Administration.   The Observatory was made possible by the generous financial support of the W.M. Keck Foundation.  We thank the Keck Observatory support staff for their assistance.  We recognize and acknowledge the very significant cultural role and reverence that the summit of Mauna Kea has always had within the indigenous Hawaiian community.  We are most fortunate to have the opportunity to conduct observations from this mountain.

\section{APPENDIX}

As noted in Section 2, the HIRES wavelength coverage is not continuous over the entire range of wavelengths discussed in this paper.   Thus, some transitions -- notably the Na I doublet near 8190 \AA\ -- fall in holes in our coverage and their absence in Tables 1-4 should not be taken to indicate that they are not present in the spectra of LDS 5606 (they clearly are present in the UVES spectra, Table 6) .   Table A1 summarizes the HIRES wavelength coverage during the three nights of observation of our program.

\section*{REFERENCES}
\begin{harvard}

\item[Absil, O., Defrere, D., Coude du Foresto, V. et al. 2013, A\&A, 555, 104]
\item[Bergin, E., Calvet, N., D'Alessio, P., \& Herczeg, G. 2003, ApJ 591, L159]
\item[Beristain, G., Edwards, S. \& Kwan, J. 2001, ApJ 551, 1037]
\item[Bertout, C. 1989, ARA\&A, 27, 351]
\item[Binks, A. \& Jeffries, R. 2014, MNRAS, in press (arXiv:1310.2613)]
\item[Cataldi, G., Brandeker, A., Olofsson, G. et al, 2014, A\&A, 563, A66]
\item[Chen, C. \& Jura, M. 2001, APJ 560, L171]
\item[Curran, R., Argiroffi, C., Sacco, G. et al 2011, A\&A 526, A104]
\item[Dekker, H., D'Odorico, S., Kaufer, A., Delabre, B. \& Kotzlowski, H. 2000, Proc. SPIE, 4008, 534]
\item[Dent, W., Wyatt, M., Roberge, A. et al 2014, Science 343, 1490]]
\item[de Val-Borro, M., Gahm, G., Stempels, H. \&  Peplinski, A. 2011, MNRAS 413, 2679]
\item[Donati, J.-F., Gregory, S., Montmerle, T. et al 2011, MNRAS 417, 1747]
\item[Ducourant, C., Teixeira, R., Galli, P. et al 2014, A\&A, in press (arXiv:1401.1935)]
\item[Gagne, J., Lafreniere, D., Doyon, R., Malo, L. \& Artigau, E. 2014, ApJ 783, 121]
\item[Gorti, U., Hollenbach, D., Najita, J. \& Pascucci, I. 2011, ApJ 735, 90]
\item[G\"usten, R., Nyman, L., Schilke, P., et al 2006, A\&A 454, L13] 
\item[Hansen, B. \& Murray, N. 2013 ApJ 775, 53]
\item[Hauschildt, P., Allard, F., \& Baron, E. 1999, ApJ 512, 377]	
\item[Hutchinson, M., Evans, A., Winkler, H. \& Spencer Jones, J. 1990, A\&A, 234, 230]
\item[Jensen, E. \& Mathieu, R. 1997, AJ 114, 301]
\item[Kastner, J., Huenemoerder, D., Schulz, N., Canizares, C. \& Weintraub, D. 2002, ApJ 567, 434]
\item[Kastner, J., Thompson, E., Montez, R., et al 2012, ApJ 747, L23]
\item[Kiefer, F., Lecavelier des Etangs, A., Augereau, J.-C. et al 2014, A\&A 561, L10]
\item[Kospal, A., Moor, A., Juhasz, A. et al 2013, ApJ 776, 77]
\item[Looper, D., Bochanski, J., Burgasser, A. et al 2010b, AJ 140, 1486]
\item[Looper, D., Mohanty, S., Bochanski, J., et al 2010a, ApJ 714, 45]
\item[Malo, L., Doyon, R., Lafreniere, D., et al 2013, ApJ 762, 88]
\item[Martin, D. C., Fanson, J., Schiminovich, D., et al 2005, ApJ 619, L1 ]
\item[Melis, C, Zuckerman, B., Rhee, J. \& Song, I. 2010, ApJ 717, L57]
\item[Muzerolle, J., Calvet, N. \& Hartmann, L. 2001, 550, 944]
\item[Natta, A., Testi, L., Muzerolle, J., et al 2004, A\&A 424, 603]
\item[Nidever, D., Marcy, G., Butler, R. P., Fischer, D. \& Vogt, S. 2002, ApJS 141, 503]
\item[Ortega, V., de la Reza, R., Jilinski, E. \& Bazzanella, B., 2002, ApJ 575, L75]
\item[Osterbrock, D. 1974, "Astrophysics of Gaseous Nebulae", San Francisco, W. H. Freeman \& Co.] 
\item[Pecaut, M \& Mamajek, E. 2013, ApJS 208, 9]
\item[Riaz, B., Gizis, J. \& Harvin, J. 2006, AJ 132, 866]
\item[Rigliaco, E., Pascucci, I., Gorti, U., Edwards, S. \& Hollenbach, D., 2013, ApJ 772, 60]
\item[Riviere-Marichalar, P., Barrado, D. Augereau J.-C. et al. 2012, A\&A 546, L8]
\item[Rodriguez, D., Kastner, J., Wilner, D. \& Qi, C., 2010, ApJ 720, 1684]
\item[Rodriguez, D., Bessell, M., Zuckerman, B. \& Kastner, J., 2011, ApJ  727, 62]
\item[Rodriguez, D., Zuckerman, B., Kastner, J., et al 2013, ApJ 774, 101]
\item[Rodriguez, D., Zuckerman, B., Faherty, J. \& Vican, L. 2014, A\&A, in press (arXiv:1404.2543]
\item[Rodriguez, D. \& Zuckerman, B. 2012, ApJ 745, 147]
\item[Rosenfeld, K., Andrews, S., Wilner, D., Kastner, J. \& McClure, M. 2013, ApJ 775, 136]
\item[Rosenfeld, K., Qi, C., Andrews, S., et al 2012, ApJ 757, 129]
\item[Scheneider, A., Melis, C. \& Song, I. 2012, ApJ 754, 39]
\item[Shkolnik, E., Liu, M., Reid, I. N., Dupuy, T., Weinberger, A. 2011, ApJ 727, 6]
\item[Skrutskie, M., Cutri, R., Stiening, R., et al 2006, AJ 131, 1163]
\item[Smits, D. 1991, MNRAS 248, 193]
\item[Stempels, H. \& Gahm, G. 2004, A\&A 421, 1159]
\item[Sytov, A., Kaigorodov, P., Fateeva, A. \& Bisikalo, D. 2011, Astr. Rep. 55, 793 ]
\item[Torres. C. et al. 2008, in Handbook of Star Forming Regions, Vol II, ed B. Riepurth, 757]
\item[Vican, L. \& Schneider, A. 2014, ApJ 780, 154]
\item[Vogt, S., Allen, S., Bigelow, B., et al. 1994, SPIE 2198, 362]
\item[White, R. \& Basri, G. 2003, ApJ 582, 1109]
\item[Wright, E., Kirkpatrick, J.D., Gelino, C., et al 2014, AJ 147, 61]
\item[Zuckerman, B. \& Song, I. 2004, AR\&A 42, 685]
\item[Zuckerman. B. \& Song, I. 2012, ApJ 758, 77] 
\item[Zuckerman, B., Song, I, Bessell, M. \& Webb, R. 2001, ApJ 562, L87]
\item[Zuckerman, B., Melis, C., Rhee, J., Schneider, A. \& Song, I. 2012, ApJ 752, 58]
\end{harvard}

\clearpage

\begin{deluxetable}{@{}lccc}
\tablewidth{0pt}
\tablecaption{HIRES Measured Emission Lines in LDS 5606A}

\tablehead{UT date & vacuum & Ion  &  EW \\
(2013) & wavelength (\AA) &  & (m\AA)}
\startdata
\hline
10/17	& 	4377.159	& Fe I]	&	96 \\
10/17	 &	4384.776	& Fe I]	&	193 \\ 
10/17	& 	4389.162	&	He I	&	~700 \\
10/17	 &	4396.267  &		Ti II	&	105 \\
10/17	& 	4405.987	&		Fe I]	&	137 \\
10/17	& 	4416.362	&		Fe I]	&	108 \\
10/17	& 	4428.553  &		Fe I]	&	200 \\
10/17	 &	4445.041	 &		Ti II	&	90 \\
10/17	& 	4469.761	 &		Ti II	&	60 \\
10/17	 &	4472.740   &		He I	&	4900 \\
10/17	 &	4502.535	&		Ti II	&	78     \\
10/17	 &	4535.240	& Ti II 	&	217   \\
 10/17	 &	4550.893  &		Ti II	&	208    \\
10/17	 &	4687.020	&		He II	 &	1360  \\
10/17	 &	4714.460  &		He I	&	500  \\
10/17	 &	4862.683   &		H I	&	90000  \\
11/16	   &	4862.683		&	H I	&	81535  \\
10/17  &	 	4923.305	&		He I	 &	863  \\
11/16	  &	4923.305		&		He I	 &	722  \\
10/17  &		4925.302	&		Fe II	 &	136  \\
11/16	   &	4925.302		&		Fe II	 &	100  \\
10/17   &		4928.192	&		Fe I] (?)  &	112  \\
11/16	   &	4928.192		&		Fe I] (?) &	97  \\
10/17   &		4928.793  &		Fe I] (?) &	72  \\
11/16	   &	4928.793		&		Fe I] (?)  &   84  \\
10/17   &		4929.707	&		Ti I (?)	&	124  \\
11/16	   &	4929.707		&	Ti I (?)	&	115  \\
10/17   &		5017.076	&		He I	&	1075  \\
11/16	   &	5017.076		&		He I	 &	988  \\
10/17	&	5019.840	&		Fe II	 &	200  \\
11/16	   &	5019.840		&		Fe II	 &	191  \\
10/17	&	5168.760	&		Mg I	 &	459  \\
11/16	   &	5168.760		&		Mg I	 &	364  \\
10/17	&	5170.473	&		Fe II 	 &	553  \\
11/16	   &	5170.473		&		Fe II 	 &	421  \\
10/17	&	5174.125	&		Mg I	 &	432  \\
11/16	  &	5174.125		&		Mg I	 &	371  \\
10/17    &		5185.048	&		Mg I	&	281 \\
11/16	   &	 5185.048		&		Mg I	 &	269  \\
10/17   &		5578.887  &		[O I]	 &	223  \\
11/16	   &	5578.887		&		[O I]	 &	171  \\
10/17   &	5877.245		&		He I	 &	8117  \\
11/16	   &	5877.245		&		He I	 &	7000  \\
10/17   &		5891.583  &		Na I	 &	1340  \\
11/16	   &	5891.583		&		Na I	 &	1000  \\
10/17	&	5897.558	&		Na I	 &	758  \\
11/16	   &	5897.558		&		Na I	 &	568  \\
10/17  &		6232.446	&		Fe I	 &	80  \\
11/16  &		6232.446	&		Fe I	 &	78  \\
10/17   &		6302.046	&		[O I]	 &	342  \\
11/16	  &	6302.046		&	[O I]	 &	306  \\
10/17    &		6564.610	&		H I	 &	99400  \\
11/16	   &	6564.610		&		H I	 &	135300  \\
10/17	&	6679.996	&		He I	 &	1773  \\
11/16	  &	6679.996		&		He I	 &	1477  \\
11/16	   &	8500.360		&		Ca II	 &	184	 \\	
11/16	   &	8544.440		&		Ca II	 &	220  \\
11/16	   &	8664.520		&		Ca II	 &	P Cyg?  \\
\enddata
\tablecomments{Rest wavelengths are from the Atomic Line List http://www.pa.uky.edu/$\sim$peter/newpage/}
\end{deluxetable}

\clearpage

\begin{table}
\caption{HIRES Measured Absorption Lines in LDS 5606A}
\begin{tabular}{@{}lccc}
\br
UT date & vacuum & Ion  &  EW \\
(2013) & wavelength (\AA) &  & (m\AA) \\
\mr
10/17	&	4417.708	&	V I	&	73(?) \\  
10/17	&	4595.411	&	V I	&	184 \\
10/17	&	4738.975	&	Sc I (?) &		247 \\
11/16	  &	4738.975	&	Sc I (?)	&	100 \\
10/17  &		4935.461	&	Fe I]	&	100 \\
11/16	  &	4935.461	&	Fe I]	 &	94 \\
10/17  &  5347.288  &  Cr I  & 139 \\
11/16     &   5347.288  &  Cr I  &  146 \\
10/17   &   5349.799  &   Cr I   &  109 \\
11/16   &   5349.799   &   Cr I  &   120 \\
10/17   &   5350.953   &   Ca I   &   119 \\
11/16   &   5350.953   &   Ca I   &  100 \\
10/17  &		5396.176	&	Mn I] (?)  &	149 \\
11/16	  &	5396.176	&	Mn I] (?) &	133 \\
10/17  &		5427.745	&		Ti I] (?)  &	66 \\
11/16	 &	5427.745	&	Ti I] (?)  &	54 \\
10/17  &		5434.056  &		Mn I] (?)  &	118 \\
11/16	  &	5434.056	&	Mn I] (?) &	127 \\
10/17  &		6123.912	&		Ca I   &		68 \\
11/16	 &	6123.912	&   Ca I	&	56 \\
10/17  &		6440.855	&		Ca I   &		104 \\
11/16	  &	6440.855	&	Ca I	&	111 \\
10/17  &		6464.353	&		Ca I   &		179 \\
11/16	  &  6464.353	&	Ca I	 &	228 \\
10/17	&	6709.640	&		Li I    &		382 \\
11/16	 &	6709.640	&	Li I	&	380 \\
10/17     &    7328.164    &   Ca I   &   279 \\
11/16	 &	7328.164	&	Ca I	&	254 \\
11/16	 &	7701.093	&				K I	&	1180 \\
11/16	 &	7802.414	&				Rb I	&	290 \\
11/16	 &	7949.789	&				Rb I	&	230  \\
11/16	 &	8049.830	&	Fe I (?)	&	156 \\
11/16	 &	8077.370	&	Fe I (?)	&	137 \\
\br
\end{tabular}
\end{table}

\clearpage

\begin{table}
\caption{HIRES Measured Emission Lines in LDS 5606B}
\begin{tabular}{@{}lcc}
\br
vacuum & Ion  &  EW \\
wavelength (\AA) &  & (m\AA) \\
\mr
4389.162	& He I		&	114 \\
4472.740	&	He I		&	1640 \\
4862.683	&			H I	&		38910 \\
5017.076	&			He I	&		534 \\
5170.473	&			Fe II    & 		472 \\
5578.887	&			[O I]		&	100 \\
5877.250	&			He I		&	3400 \\
5891.583	&			Na I		&	1540 \\
5897.558	&			Na I		&	827 \\
6302.046	&			[O I]		&	93 (?) \\
6564.610	&			H I		&	25100 \\
6679.996	&			He I		&	777 \\
\br
\end{tabular}
\end{table}

\clearpage

\begin{deluxetable}{@{}lcc}
\tablewidth{0pt}
\tablecaption{HIRES Measured Absorption Lines in LDS 5606B}
\tablehead{vacuum & Ion  &  EW \\
wavelength (\AA) &  & (m\AA) }
\startdata
\hline
4413.474	& 	Cr I		&	294 \\
4417.708	&		V I	&		128  \\
4436.924	&			Ca I    &			374 \\
4458.680	&		Ti I (a)		&	125 \\
4460.980	&			Cr I		&	120 \\
4461.400	&			Fe I]		&	196 \\
4595.410	&			V I		&	237 \\
4608.622	&			Sr I		&	802 \\
5298.165	&			Cr I		&	166 \\
5299.751	&			Cr I		&	172 \\
5329.521	&			Fe I		&	~500 \\
5342.509	&			Fe I		&	146 \\
5347.288	&			Cr I		&	224 \\
5349.799	&			Cr I		&	225 \\
5350.953	&			Ca I		&	148 \\
5372.983	&			Fe I		&	216 \\
5396.176	&			Mn I]		&	213 \\
5398.628	&			Fe I		&	313 \\
5407.277	&			Fe I		&	224 \\
5411.276	&			Cr I		&	177 \\
5427.745	&			Ti I] (?)	&	86 \\
5431.206	&			Fe I		&	230 \\
5434.056	&			Mn I] 	&		178 \\
5590.301	&			Ca I		&	147 \\
5591.666	&			Ca I		&	170 \\
5596.015	&			Ca I		&	101 \\
5689.783	&			Na I		&	145 \\
6104.412	&			Ca I		&	111  \\
6123.912	&			Ca I		&	275  \\
6163.878	&			Ca I		&	114 \\
6440.855	&			Ca I		&	162 \\
6464.353	&			Ca I (b)	&	226 \\
6473.450	&			Ca I		&	127(?) \\
6709.640	&			Li I		&	501  \\
7701.093	&			K I		&	1360 \\
7802.414	&			Rb I		&	355 \\
7949.789	&			Rb I		&	268  \\
8049.830	&			Fe I 		&	215  \\
8077.370	&			Fe I 		&	158  \\
\enddata
\tablecomments{
Rest wavelength for $\lambda$ 4608.622 Sr I line is from 
http://physics.nist.gov/PhysRefData/ASD/lines$\_$form.html
Notes: (a) possible contribution from Mn I at wavelength  4458.800 \AA.
(b) possible contribution from Fe I at 6464.512 \AA.
}
\end{deluxetable}

\clearpage

\begin{table}
\caption{UVES Measured Emission Line Flux Densities}
\begin{tabular}{@{}lcc}
\br
star & transition & flux    \\
  &  &  density \\
\mr
A & H-11 & 76 \\
A  &  H-10  &   101 \\
A  &  H$\eta$   & 144 \\
B &  H$\zeta$  &  16 \\
A  &  H$\zeta$  &  190 \\
B  &   H$\epsilon$  & 15 \\
A  & H$\epsilon$   &  291 \\
B  &  H$\delta$  &  28 \\
A &  H$\delta$    &  289 \\
B  &  H$\gamma$  &  44  \\
A &  H$\gamma$  &  403  \\
B  & H$\beta$  &  80 \\
A  &  H$\beta$  &  503  \\
B  &  Ca II K  &  37 \\
A &  Ca II K  &  50  \\
B  &  Ca II H  &  26 \\
A &  Ca II H  &  27 \\
\br
\end{tabular}
\end{table}
Note:  A and B stand for LDS 5606A and B.  The flux density units are 10$^{-16}$ erg cm$^{-2}$ s$^{-1}$.  See also Figures 7 and 8..

\clearpage

\begin{table}
\caption{UVES Measured Lines in LDS 6506}
\begin{tabular}{@{}lccc}
\br
star &  vacuum & Ion  &  EW \\
&  wavelength (\AA) &  & (\AA) \\
\mr
B & 6302.046 & [O I] & -0.160 \\
A & 6302.046 & [O I] & -0.470 \\
B & 6564.610 &  H I  &  -16.0 \\
A & 6564.610  &  H I  &  -84.9 \\
B &  6679.996 & He I  & -0.378 (?)  \\
A &  6679.996 &  He I  &  -1.939 \\
B & 6709.640  & Li I  & 0.489 \\
A  & 6709.640 &  Li I &  0.383 \\
B &  7067.127  &  He I  &   $-$ \\
A &  7067.127  &  He I  &  -0.600 \\
B &  7701.093  &  K I  & 1.447 \\
A &  7701.093  &  K I  &  1.250 \\
B  & 7802.414  &  Rb I  &  0.309 \\
A  &  7802.414  &  Rb I  &  0.303  \\
B  &  7949.789  &  Rb I  &  0.270 \\
A  &  7949.789  &  Rb I  &  0.233 \\
B  &  8049.830  &  Fe I  &  0.208 \\
A  &  8049.830  &  Fe I  &  0.171 \\
B  &  8077.370  &  Fe I  &  0.150 \\
A  &  8077.370  &  Fe I  &  0.167 \\
B  &  8185.505  &  Na I  &  1.550  \\
A &  8185.505  &  Na I  &  1.573  \\
B  &  8197.077  &  Na I  &  2.060  \\
A  &  8197.077  &  Na I  &  2.183  \\

\br
\end{tabular}
\end{table}
Note:  A and B stand for LDS 5606A and B.  Negative entries indicate emission lines.
For the B star, the upper limit to the EW of a helium emission line at 7067.1 \AA\ is $\sim$80 m\AA.

\clearpage

\begin{table}
\caption{Veiling in LDS 5606}
\begin{tabular}{@{}lcccc}
\br
star & range & range & range & range   \\
&  4400$-$4600 \AA\ & 5300$-$5700 \AA\ &  6100$-$6700 \AA\ &  7700$-$8100 \AA\ \\
\mr
LDS 5606B & 0.14 & 0.5 & 0.8 & 0 \\
LDS 5606A & 1.5 & 2.7 & 1.9 & 0.3 \\
\br
\end{tabular}
\end{table}
Note:  the table entries are r = F$_{excess}$/F$_{photosphere}$ (see Section 4.1).

\clearpage

\begin{table}
\caption{LDS 5606A: Photospheric Absorption \\ and Helium \& Oxygen Emission Line Velocities}
\begin{tabular}{@{}lcccc}
\br
& Ion  & wavelength & redshift &  redshift \\
& & (\AA) &  (\AA) &  (km s$^{-1}$) \\
\mr
& Fe I]	 & 4935		&	0.254	&		15.4 \\
&Mn I]	  & 5396		&	0.269	&		15 \\
&Mn I]	 & 5434		&	0.244	&		13.5 \\
&Ca I	& 6441		&	0.285	&		13.3 \\
&Ca I	&  6464		&	0.30		&	14 \\
&Li I		& 6710		&	0.38		&	17 \\
&Ca I	&  7328		&	0.346	&		 14 \\
&Fe I	&  8050		&	0.38		&	14 \\
&Fe I	&  8077		&	0.39		&	14.5 \\
&He I	&  4389		&	0.33		&	23  \\
&He I	&  4473		&	0.33		&	22 \\
&He I	&  4714		&	0.36		&	23 \\
&He I	&  4923		&	0.33		&	20 \\
&He I	&  5017		&	0.29		&	17  \\
&He I	&  5877		&	0.43		&	22  \\
&He I	&  6680		&	0.44		&	20  \\
&He II	&  4687		&	0.49		&	31  \\
& [O I]	&  5579		&	0.31		&	16.6  \\
& [O I]	&  6302		&	0.29		&	14  \\
\br
\end{tabular}
\end{table}
Notes:  The mean of the heliocentric velocities of the 9 photospheric absorption lines (from elements not helium or oxygen) is 14.5$\pm$0.4 km s$^{-1}$.  The mean velocity of the 7 emission lines from He I is 21$\pm$0.9 km s$^{-1}$.  The velocity of the one emission line from He II is 31 km s$^{-1}$.  The mean velocity of the two forbidden oxygen emission lines is 15.3 km s$^{-1}$.

\clearpage

\setcounter{table}{0}
\renewcommand{\thetable}{A\arabic{table}}

\begin{table}
\caption{HIRES Wavelength Coverage (\AA)}
\begin{tabular}{@{}lccc}
\br
CCD & 10/17/13 & 10/21/13  &  11/16/13 \\
& Begin \ \ \ \ \ End & Begin \ \ \ \ \ End & Begin \ \ \ \ \ End \\
\mr
1 &	4364	\ \ \ \	5744	 &  4364 \ \ \ \  5744 &	4717	\ \ \ \  5986 \\
2 &	5823	\ \ \ \	6540	&   5822 \ \ \ \	6540 & 6067 \ \ \ \  6545 \\
2 &	6545	\ \ \ \	6661&  6546 \ \ \ \	6661 & 6550 \ \ \ \  6666 \\
2 &	6669	\ \ \ \	6787	 &  6669 \ \ \ \	6787 & 6675 \ \ \ \  6791 \\
2 &	6797	\ \ \ \	6917	  & 6797 \ \ \ \	6917	& 6803 \ \ \ \  6921 \\
2 &	6931	\ \ \ \	7053	 &  6931 \ \ \ \	7052	& 6937 \ \ \ \  7054 \\
2 &	7069	\ \ \ \	7194	 &  7070 \ \ \ \	7194	& 7075 \ \ \ \  7198 \\
2 &	7214	\ \ \ \	7328	 &  7214 \ \ \ \	7328	& 7220 \ \ \ \  7345 \\
2 &				&				&	7370	\ \ \ \  7498 \\
2 &				&				&	7527	\ \ \ \ 7583 \\
3 &	7428\ \ \ \	7496	  &  7425 \ \ \ \	7494 & 7700\ \ \ \ 7834 \\
3 &	7521\ \ \ \ 7653	   & 7521 \ \ \ \	7653 & 7865\ \ \ \  8004 \\
3 &	7684\ \ \ \ 7819	   & 7684 \ \ \ \	7819 & 8039\ \ \ \  8181 \\
3 &	7855\ \ \ \ 7993	   & 7855 \ \ \ \	7993 & 8227\ \ \ \ 8371 \\
3 &	8034\ \ \ \ 8175	   & 8034 \ \ \ \	8174 & 8423\ \ \ \ 8570 \\
3 &	8221\ \ \ \ 8366	  & 8220 \ \ \ \	8365 & 8628\ \ \ \ 8778 \\
3 &   8417\ \ \ \ 8565  &     				& 8844\ \ \ \ 9000 \\
3 &    8621\ \ \ \ 8773  &    &  \\
\br
\end{tabular}
\end{table}
\noindent Note $-$  The UVES wavelength coverage was 3762 to 4986 \AA, 5692 to 7526 \AA, and 7662 to 9460 \AA.

\clearpage

\begin{figure}
\includegraphics[width=200mm]{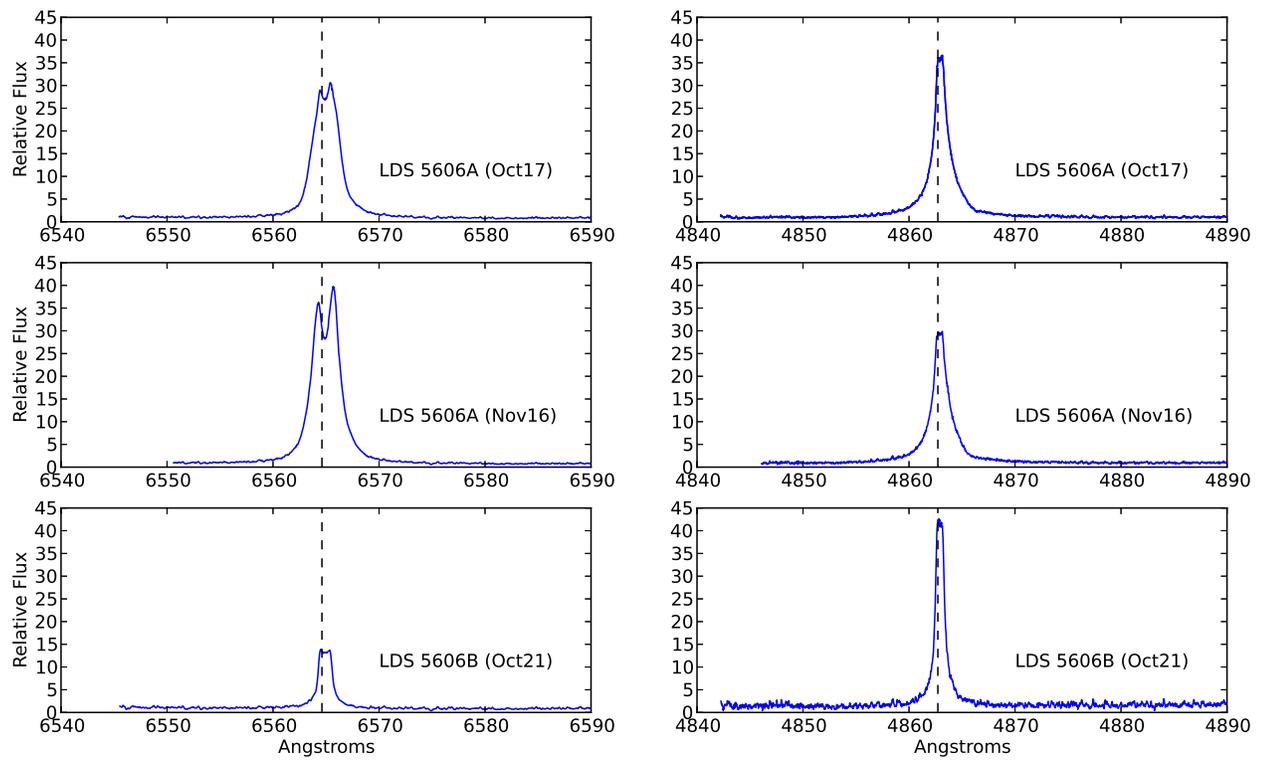}
\caption{\label{figure1} H$\alpha$ and H$\beta$ lines in LDS 5606A and LDS 5606B.  The abscissa is in vacuum in the heliocentric rest frame.  The vertical dashed lines indicate zero velocity.}
\end{figure}

\clearpage

\begin{figure}
\includegraphics[width=200mm]{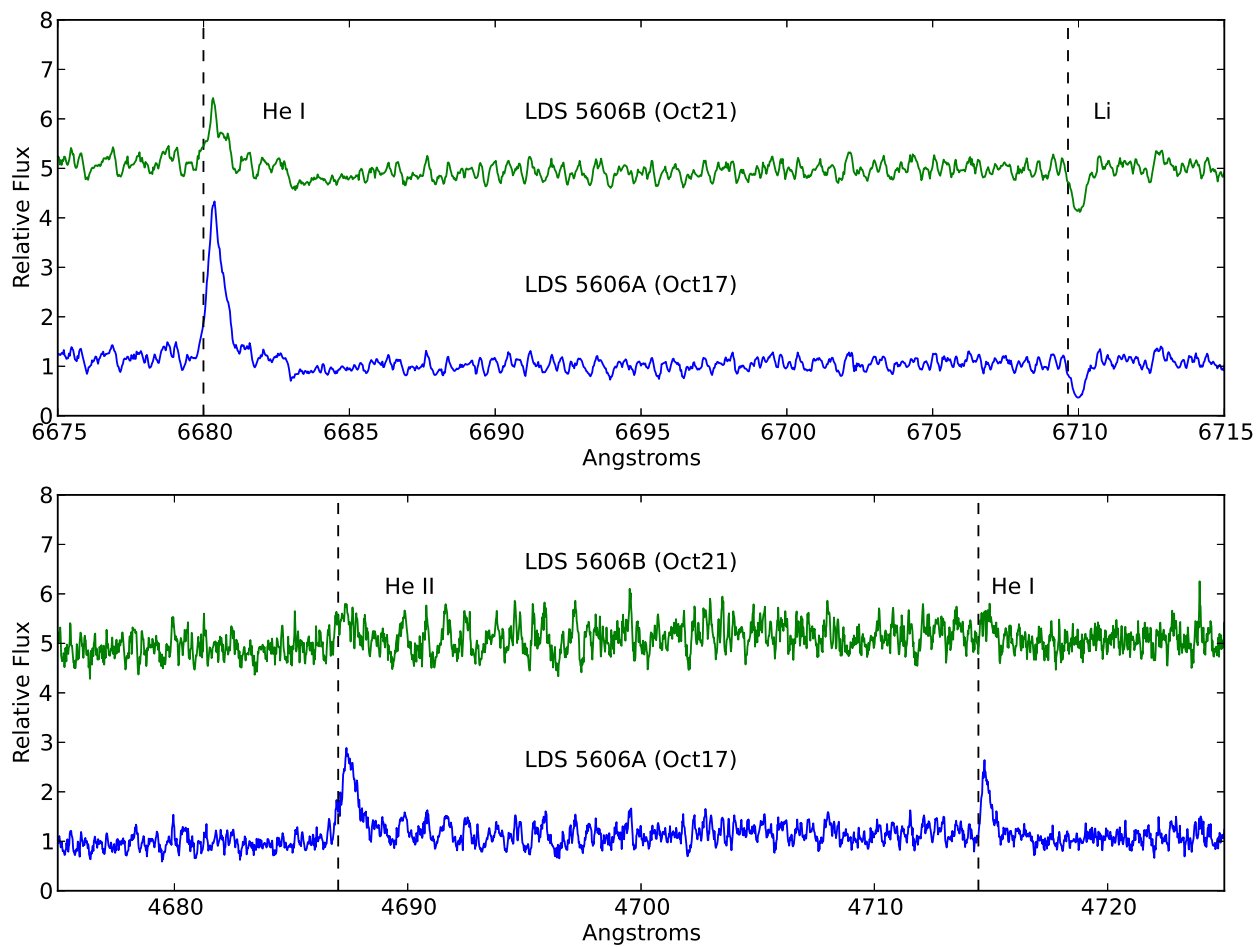}
\caption{\label{figure2}$\it{Upper~Panel:}$ Transitions of He I and Li I.  $\it{Lower~Panel:}$ An He I line and the Paschen $\beta$ line of He II.  A possible detection of the latter line in LDS 5606B requires confirmation. The abscissa and vertical dashed lines are as in Fig. 1.}
\end{figure}

\clearpage

\begin{figure}
\includegraphics[width=200mm]{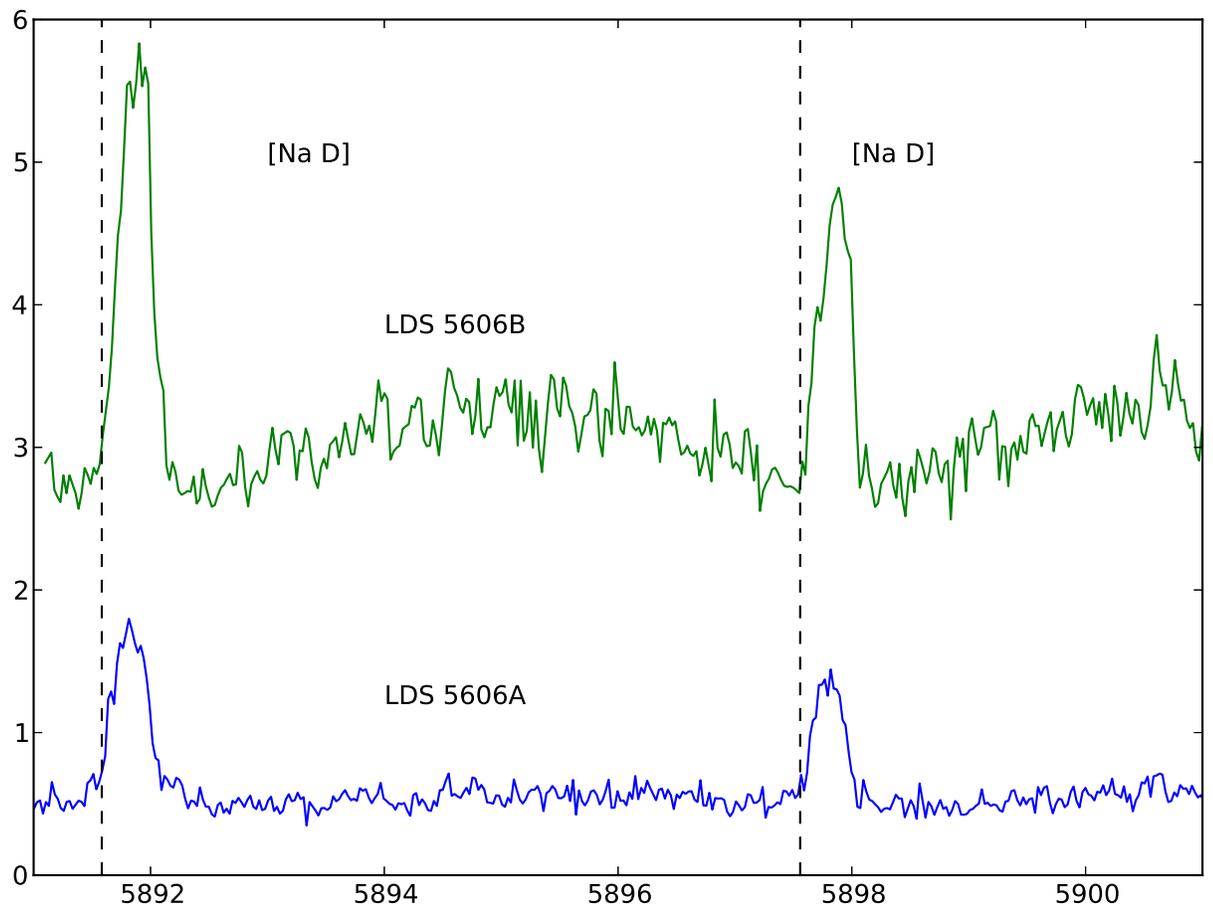}
\caption{\label{figure3} The Na D lines.  Veiling of the absorption line component in LDS 5606A is apparent. The abscissa and vertical dashed lines are as in Fig. 1.}
\end{figure}

\clearpage

\begin{figure}
\includegraphics[width=200mm]{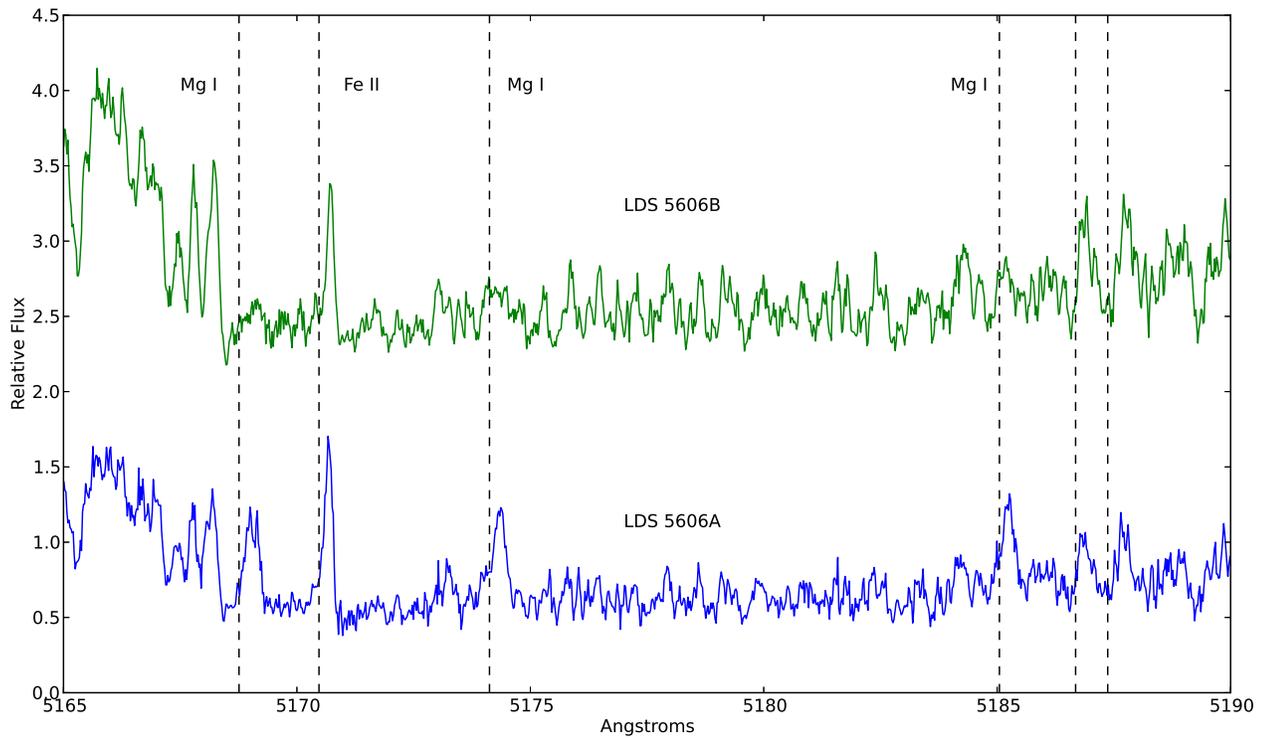}
\caption{\label{figure4}Representative emission lines.  The complex structure of the spectrum blueward of the 5168.7 \AA\ line of Mg I is a TiO bandhead. The abscissa and vertical dashed lines are as in Fig. 1.}
\end{figure}

\clearpage

\begin{figure}
\includegraphics[width=200mm]{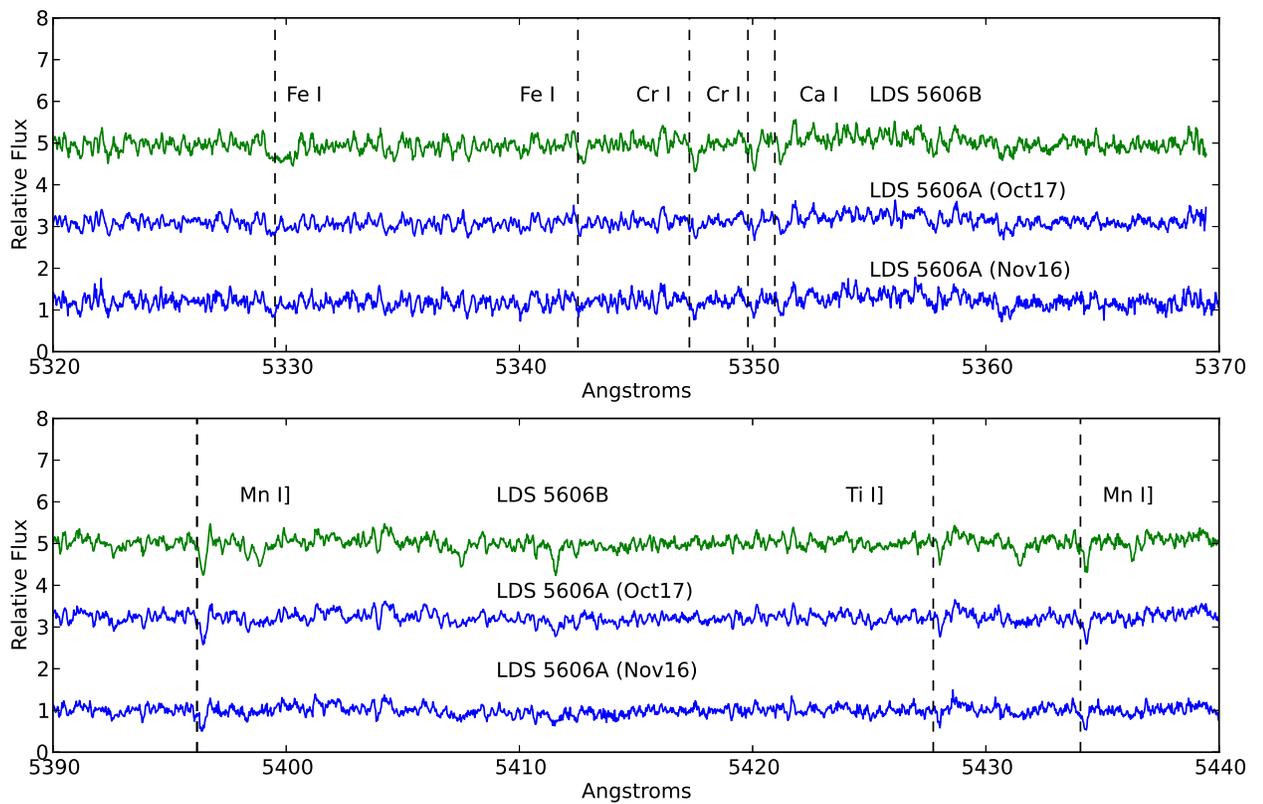}
\caption{\label{figure5} Representative absorption lines. The abscissa and vertical dashed lines are as in Fig. 1.}
\end{figure}

\clearpage

\begin{figure}
\includegraphics[width=200mm]{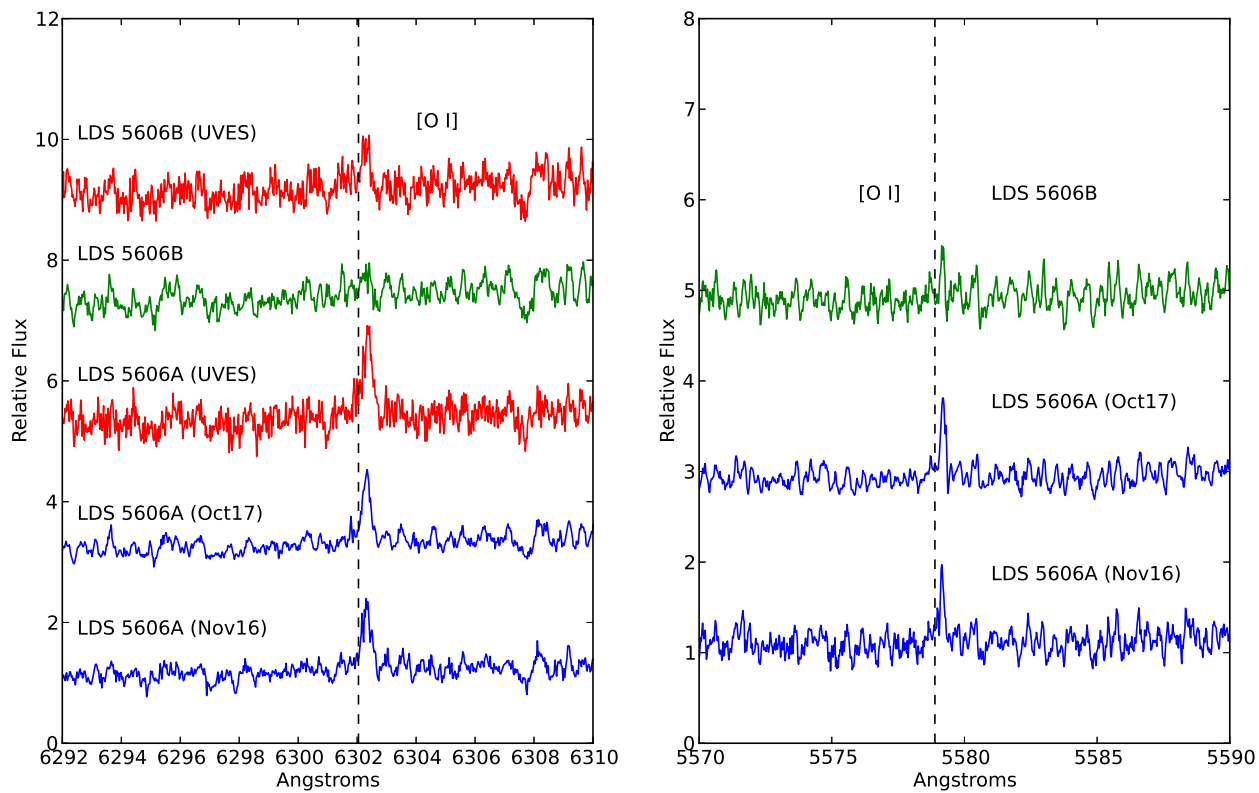}
\caption{\label{figure6}$\it{Left~Panel:}$ $^1$D$_2$$-$$^3$P$_2$ forbidden oxygen line from HIRES and UVES. $\it{Right~Panel:}$ $^1$S$_0$$-$$^1$D$_2$ forbidden oxygen line from HIRES. All spectra not labelled as (UVES) are from HIRES.  The abscissa and vertical dashed lines are as in Fig. 1.}
\end{figure}

\clearpage

\begin{figure}
\includegraphics[width=200mm]{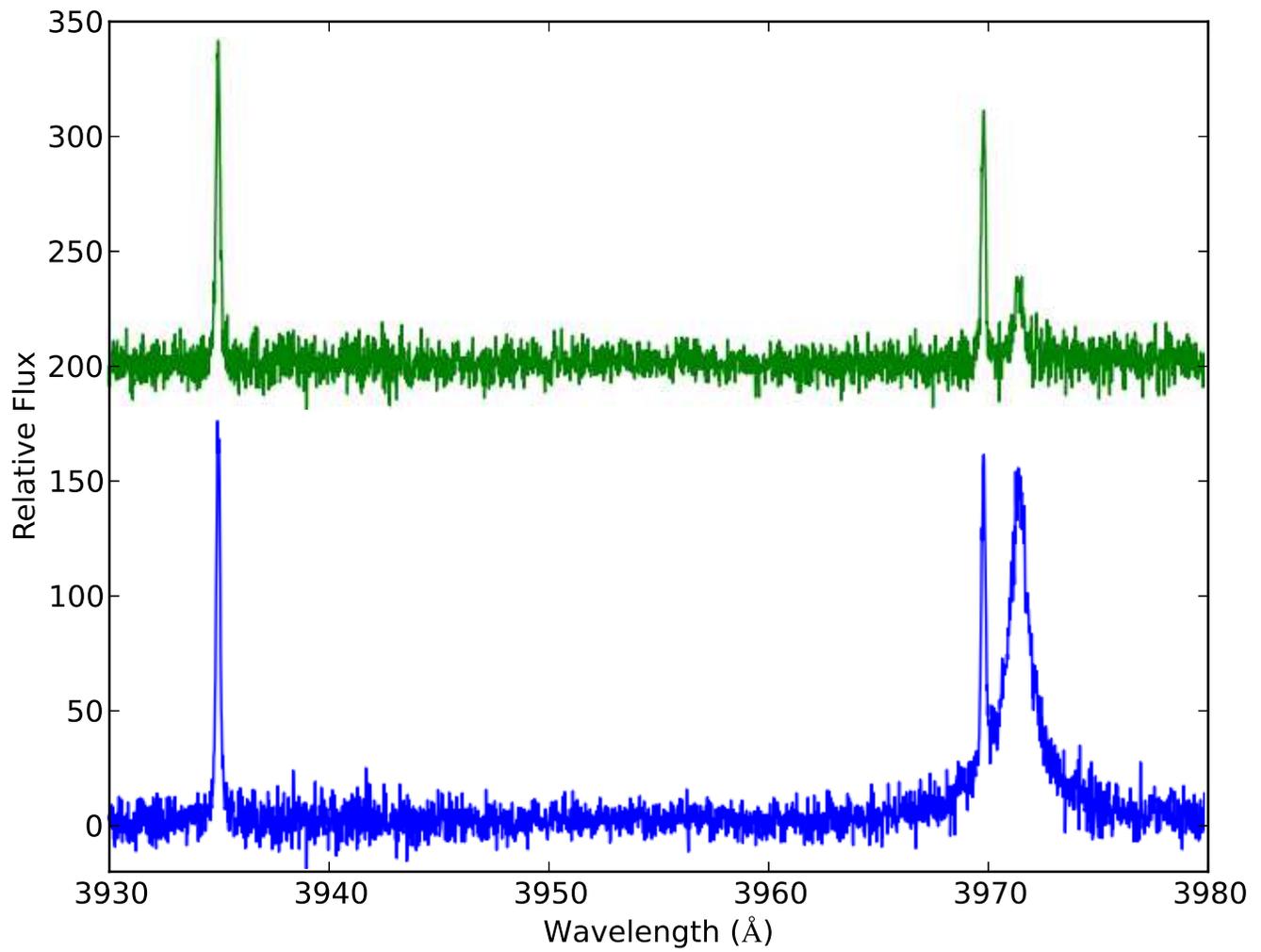}
\caption{\label{figure7}  Ca II H- and K-lines and H$\epsilon$ line in LDS 5606B (top) and LDS 5606A (bottom) from UVES spectra. The abscissa is as in Fig. 1.}
\end{figure}

\clearpage

\begin{figure}
\includegraphics[width=200mm]{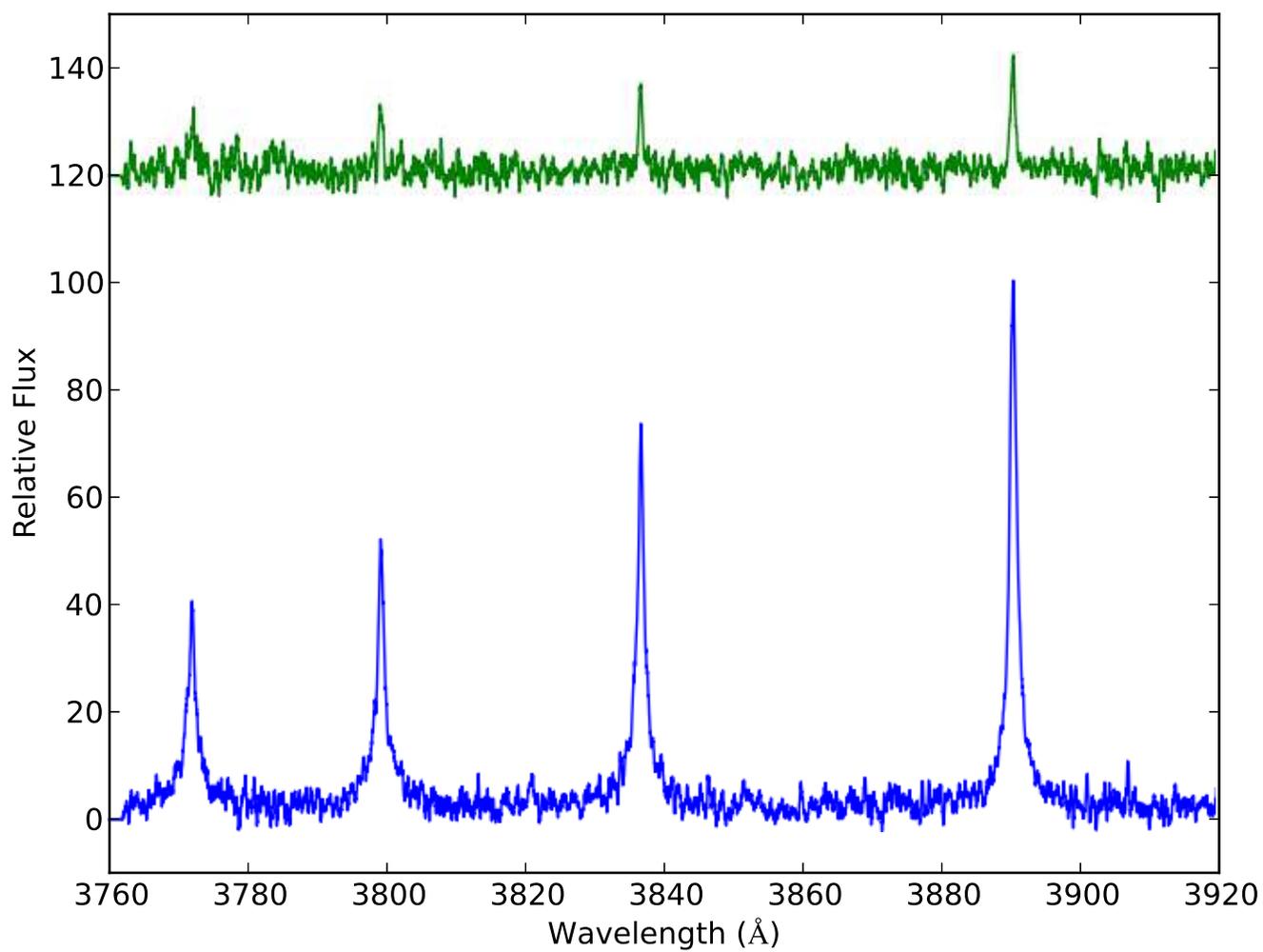}
\caption{\label{figure8} High order Balmer lines in LDS 5606B (top) and LDS 5606A (bottom) from UVES spectra.  The spectra have been smoothed by 21 pixels (0.3 \AA) to increase the signal-to-noise ratio. The abscissa is as in Fig. 1.}
\end{figure}

\clearpage

\begin{figure}
\includegraphics[width=150mm]{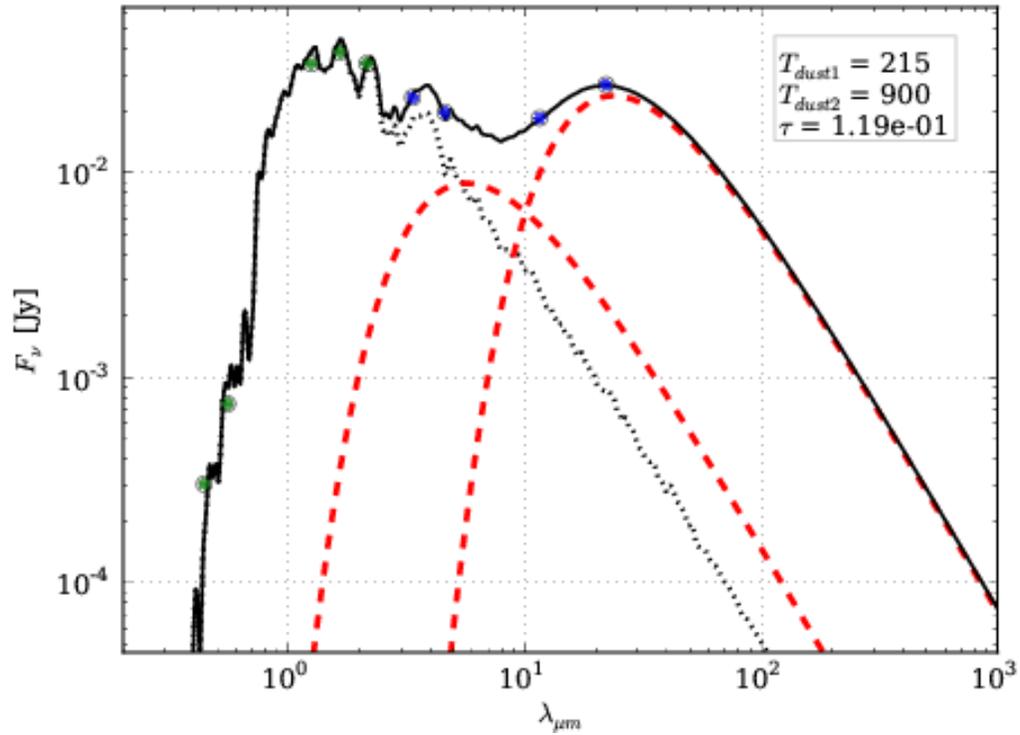}
\caption{\label{figure9}  SED of LDS 5606A.  The solid black curve is a fit of a Hauschildt et al (1999) photospheric model (see Vican \& Schneider 2014 for details) plus blackbodies with temperatures of 215 and 900 K.  The blackbodies are indicated with dashed curves and the model photosphere with a dotted line.   The two green points at visual wavelengths are B and V from the GSC 2.3.2.   The three green near-IR points are from 2MASS and the four blue points are from ALLWISE.  Errors in the flux densities of the two shorter wavelength ALLWISE data points are smaller than the size of the blue points.  Tau is equal to L$_{IR}$/L$_{bol}$. Tau of the 900 K component is about 7.2\% and of the 215 K component about 4.7\% }
\end{figure}

\clearpage

\begin{figure}
\includegraphics[width=150mm]{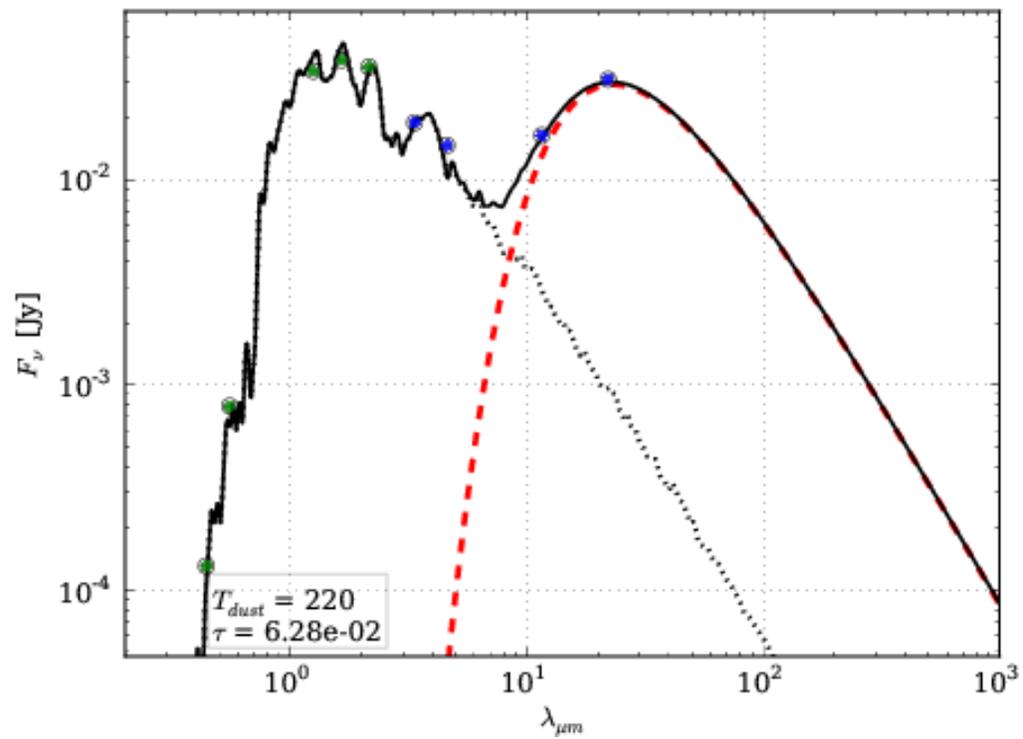}
\caption{\label{figure10} SED of LDS 5606B with a blackbody fit (dashed curve) of 220 K to the infrared excess emission, otherwise the same as Figure 9. }
\end{figure}

\clearpage

\begin{figure}
\includegraphics[width=200mm]{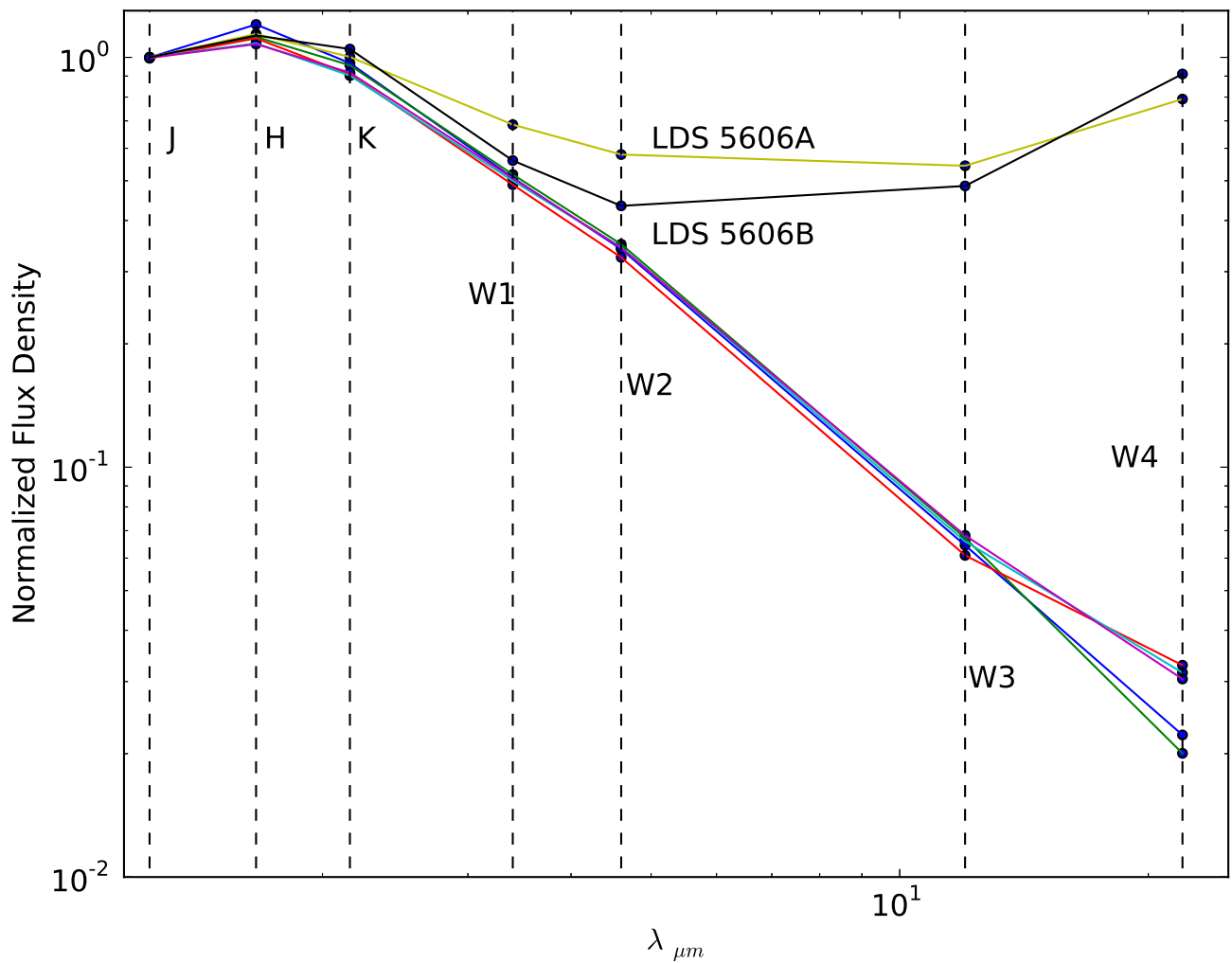}
\caption{\label{figure11}  Near to mid-IR spectra of LDS 5606A and B and five additional stars from our HIRES survey of UV-bright late-type stars (L. Vican et al, in preparation). The JHK points are from 2MASS and W1, W2, W3, and W4 are from ALLWISE.  The spectra are all normalized to the J-band flux density.  The five non-LDS 5606 stars all have spectral types in the range M4-M5 and, typically, a few H and He lines in emission.  One of the five is 2M0529-32 considered in Section 4.1.  The errors in the plotted W1 and W2 flux densities are smaller than the size of the black points.  The excess emission of LDS 5606A in W1 and W2 is evident.  }
\end{figure}

\clearpage

\begin{figure}
\includegraphics[width=200mm]{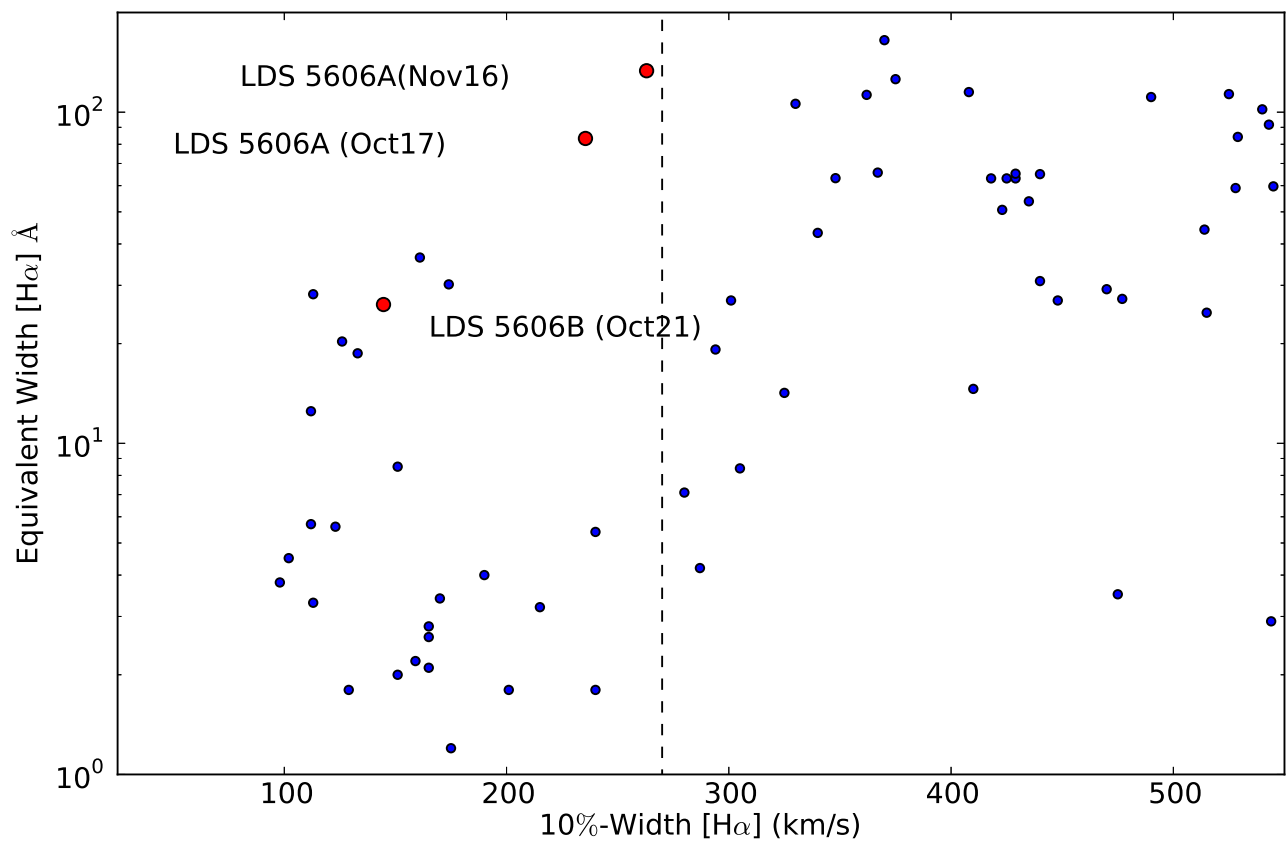}
\caption{\label{figure12}  H$\alpha$ data for the two LDS 5606 stars (large red dots) overplotted on Figure 7 of White \& Basri (2003).    The non-LDS 5606 stars plotted to the left of the vertical dashed line are weak-line T Tauri stars and those to the right of the line, with the exception of one dot (UX Tau), are classical T Tauri stars undergoing accretion. }
\end{figure}

\end{document}